\documentclass[]{jfm}

\usepackage{graphicx}
\usepackage{newtxtext}
\usepackage{comment}
\usepackage{newtxmath}
\usepackage{natbib}
\usepackage{hyperref}
\hypersetup{
    colorlinks = true,
    urlcolor   = blue,
    citecolor  = black,
}
\usepackage{soul}
\usepackage{array}
\usepackage[table]{xcolor}

\usepackage{lipsum}
\title{Fate of Secondary Droplets Produced by High-speed Raindrops Interacting with a Liquid Pool}

\author{Han-Hsiang Kuo\aff{1}
  \and Xuanting Hao\aff{1,}\aff{2}
  \corresp{\email{x3hao@ucsd.edu}}
 }

\affiliation{\aff{1}Mechanical and Aerospace Engineering, University of California San Diego, La Jolla, CA 92093, USA\aff{2}Scripps Institution of Oceanography, University of California San Diego, La Jolla, CA 92093, USA
}

\begin{document}
\maketitle

\begin{abstract}
\textcolor{black}{Secondary droplets produced by interactions between falling fluid drops and a liquid pool play a significant role in engineering applications and geophysical processes in nature. This study uses direct numerical simulations to investigate the dynamics of secondary droplets generated by raindrop-liquid pool interactions. \textcolor{black}{The raindrop parameters feature a realistic speed of $\sim 7\ \mathrm{m/s}$, effective diameters of $\sim 1$–$4\ \mathrm{mm}$, and surface tension values ranging from $25\%$ to twice the typical air–water interface value. The numerical configurations include both a single raindrop and two raindrops separated by distances between two and four times the raindrop diameter.} The secondary droplet size distribution, $N_d$, is found to scale with the droplet radius, $r_s$, as $N_d(r_s)\propto r_s^{-5/2}$, with additional dependencies on surface tension and raindrop diameter. When normalized according to this new scaling law, the droplet size distribution obtained from simulations with different parameter values collapses onto a single curve. Analysis of the impact morphology reveals distinct stages of raindrop interactions and identifies the formation and breakup of a central liquid film. Spatial and temporal analyses of the secondary droplets show that raindrop interaction can influence both the percentage of droplets captured by the cavity and the duration over which they re-merge with the pool. These behaviors arise from the combined effects of differences in the birth times of secondary droplets of various sizes and aerodynamic forcing associated with the cavity airflow.}

\end{abstract}

\section{Introduction}

The study of fluid drop impacts on a liquid surface is fundamental to engineering applications such as spray cooling and inkjet printing~\citep{liang2016review}.  Furthermore, similar processes in nature, particularly rainfall on the ocean, play an important role in influencing ocean surface salinity and temperature~\citep{li2016north, shackelford2022rain, balaguru2022impact}. Ocean raindrops can also bias rainfall measurements, as instruments may overestimate rainfall intensity by detecting secondary droplets and sprays~\citep{okachi2020characteristics}, contribute to oil spill dispersion~\citep{wu2017effects}, and generate acoustic signals that can be used to monitor precipitation patterns~\citep{prosperetti1993impact}.

\citet{worthington1883impact} pioneered the visualization of drop--surface interactions, laying the foundation for the extensive research that followed. Many configurations have since been studied, including impacts on both solid and liquid surfaces, \textcolor{black}{as well as gas film-mediated dynamics~\citep{alventosa2023inertio, sprittles2024gas}.} For liquid surfaces, the dynamics differ markedly between shallow and deep pools: shallow pools exhibit confined cavity formation and suppressed splashing due to proximity to the solid base~\citep{li2019characteristics,tang2019bouncing, wu2022droplet}, whereas deep pools permit larger cavities, higher jets, and more vigorous splashing~\citep{sykes2023droplet}.  \citet{murphy2015splash} categorized previous studies on drop impact on a deep pool into five regimes based on their Froude number $Fr = U_0^2/(g d)$ and Weber number $We = \rho_l U_0^2 d/\sigma$, where $U_0$ is the drop impact velocity, $g$ is the gravity, $d$ is the drop diameter, $\rho_l$ is the liquid density, and $\sigma$ the surface tension. The first regime, coalescence and vortex ring (C\&VR, $40<Fr<100$, $We<100$), involves drops that fall with negligible momentum and coalesce with the liquid, where high surface tension results in the formation of a downward-moving vortex ring after impact\textcolor{black}{~\citep{hsiao1988critical,deng2007role, dong2023pinching, anirudh2024coalescence, behera2026effect}}. The second regime, swell and thin jet (S\&TJ, $100<Fr<500$, $100<We<300$), is characterized by the formation of an impact cavity and outward swell; the cavity may retract and form a rising liquid jet that disintegrates into secondary droplets~\citep{manzello2002experimental, berberovic2009drop, castillo2015droplet}. The regular entrainment (RE) regime is a sub-regime of S\&TJ, where capillary waves travel down the cavity surface and pinch off an air bubble~\citep{elmore2001cavity,leng2001splash}. At higher energies, the crown and thick jet regime (C\&TJ, $100<Fr<1000$, $300<We<2000$) features increased cavity depth, while the outward swell rises to form a crown, from which ligaments form and disintegrate into secondary droplets. The rebounding cavity also forms a thick rising liquid jet, and the initial impact generates a cloud of small drops~\citep{bisighini2010crater,castillo2015droplet}. The most energetic regime, bubble canopy (BC, $300<Fr<7000$, $2000<We<20000$), produces large cavities and crowns; the crown rim rises and expands radially before converging to form a canopy and a liquid jet that travels in both vertical directions. The downward traveling jet can penetrate the cavity bottom and entrain bubbles into the liquid pool~\citep{murphy2015splash, sochan2018shape, wang2023analysis}. Previous studies on multi-drop impact on pools have largely focused on the C\&VR and S\&TJ regimes~\citep{santini2017experimental,guilizzoni2019synchronized,guilizzoni2022crater,poureslami2023simultaneous,wang2024impact}. In comparison, very little attention has been given to multi-drop interactions in the BC regime, where the impact morphology becomes highly complex and large numbers of secondary droplets are produced within a very short period. Since raindrops fall in the BC regime~\citep{gunn1949terminal,murphy2015splash}, it is important to understand how neighboring raindrop impacts interact and how these interactions influence the generation, size distribution, and dynamics of secondary droplets. 

Secondary droplets generated during drop impacts originate from several mechanisms. Prompt splash, which results from instabilities in the initial ejecta sheet, produces very fine droplets within a very short time before the drop fully merges with the pool~\citep{deegan2007complexities, zhang2012evolution, marcotte2019ejecta}. Crown rim breakup occurs later as rim ligaments fragment into medium and large droplets~\citep{villermaux2004ligament, eggers2008physics, murphy2015splash}. In multi-drop impacts, a distinguishing feature is the central liquid film, a vertically rising sheet that forms between adjacent droplets upon impact. Previous studies have shown that this film can reduce the splashing threshold and provide an additional pathway for secondary droplet formation~\citep{liang2018simultaneous, fest2021multiple, zhou2024numerical}. \textcolor{black}{Studies of drop impact on solid substrates have also examined secondary droplet dynamics and interaction effects, including analytical models for the diameters and velocities of ejected droplets~\citep{riboux2015diameters}, data-driven analyses of splashing thresholds~\citep{pierzyna2021data}, quantification of atomization in prompt splash and crown rim breakup~\citep{burzynski2020splashing}, and interactions involving multiple impacting drops and non-isolated impacts on substrates by~\citet{roisman2002multiple, wang2018non}.} In the context of secondary droplet formation during high-speed impacts representative of rainfall, \citet{wang2023analysis} examined the droplet population up to $4$ ms after impact and emphasized the generation of very small droplets associated with the prompt splash. \citet{liu2024experimental} recreated rainfall conditions in the laboratory and measured the secondary droplet number density at various heights above the surface. However, the highly dynamic nature of high-speed impacts has made it difficult to quantitatively determine the mechanisms responsible for secondary droplet formation, leaving a substantial gap in current understanding~\citep{veron2015ocean}.

Simulation-based studies on drop--pool interactions in the literature focus their efforts on resolving the interfacial dynamics between different phases. \citet{oguz1990bubble} used a boundary integral model to simulate drop impact, which included surface tension effects at the interface through curvature, with stabilization to keep the surface smooth. However, this method assumes irrotational flow, which fails to predict vortex rings in the pool. \citet{morton2000investigation} addressed this by utilizing the continuum surface force model proposed by~\citet{brackbill1992continuum}, where surface tension is added as a body force in the momentum equation, with the force direction given by the interface normal and its strength determined by the interface curvature. In terms of simulation construction, two-dimensional (2D) simulations on a planar domain~\citep{coppola2011insights, kharmiani2016simulation} significantly reduce computational costs, but provide only partial insight into the drop impact process in the radial direction. Additionally, it physically represents an infinite cylinder instead of a spherical drop. To address this, 2D axisymmetric simulations are more commonly employed while still keeping computational costs low\textcolor{black}{~\citep{fudge2021dipping,fudge2023drop,alventosa2023inertio, sanjay2023drop,wang2024impact,wang2025splashing}}. To capture non-axisymmetric features, three-dimensional (3D) simulations have been performed in a computational domain that is restricted to half or a quarter of the drop geometry, where the symmetric boundary condition is  applied~\citep{brambilla2015assessment,guo2017high,guilizzoni2019synchronized, guilizzoni2022crater}. Additionally, 3D simulations often employ adaptive mesh refinement to reduce the overall mesh count while preserving sufficient resolution, as demonstrated by~\textcolor{black}{\citet{cimpeanu2018three, liang2018simultaneous,liang2019successive, wang2023analysis}}. Since axisymmetric 2D simulations or 3D simulations with symmetric domains may exclude transverse instabilities in the crown and the resulting secondary droplet shedding~\citep{wang2025splashing}, which are crucial to the overall impact dynamics, a fully 3D simulation is more suitable for investigating the impact morphology and secondary droplet formation.

The present study addresses these gaps by constructing a fully 3D numerical simulation of high-speed raindrops impacting a liquid pool to represent rainfall on the ocean. A dimensional analysis is performed to quantify the number density of secondary droplets as a function of the droplet radius, with the aim of providing insight into the key factors governing the droplet generation process. The impact morphology of raindrops is examined, and the stages of their interaction identified and compared across multiple inter-raindrop spacings. The spatial and temporal distributions of secondary droplets are then analyzed to reveal how the fate of these droplets is shaped by raindrop interactions. The remainder of this paper is organized as follows. The numerical solver, simulation setup, and validation are described in~\S~\ref{sec:method}. In~\S~\ref{sec:results}, we present the results and discussion, including early-time secondary droplet size distributions, impact morphology, and the influence of raindrops on secondary droplet evolution. Conclusions are given in \S~\ref{sec:conclusion}.

\section{Methodology}
\label{sec:method}
\subsection{Numerical solver}
The current study carries out DNS of the incompressible air--water system using the open-source solver Basilisk~\citep{popinet2018numerical}. In Basilisk, the volume of fluid method is used to track the evolution of the interface by following the split advection method~\citep{weymouth2010conservative} to solve the advection equation
\begin{equation}
\frac{\partial f}{\partial t} + \nabla \cdot (\boldsymbol{u} f) = 0,
\label{eq1}
\end{equation}
where $f$ represents the volume fraction of water in each cell. The Navier--Stokes equations with surface tension are
\begin{align}
&\nabla \cdot \boldsymbol{u} = 0, \nonumber\\
&\rho \frac{\partial \boldsymbol{u}}{\partial t} + \rho (\boldsymbol{u} \cdot \nabla) \boldsymbol{u} 
 = -\nabla p + \nabla \cdot [\mu (\nabla \boldsymbol{u}+\nabla \boldsymbol{u}^T)] 
 + \rho \boldsymbol{g} + \sigma \kappa \delta_s \boldsymbol{n}, \label{eq3}
\end{align}
where $\boldsymbol{u}$ is the velocity vector, $\rho=f\rho_{w}+(1-f)\rho_{a}$ is the density, $\mu=f\mu_{w}+(1-f)\mu_{a}$ is the dynamic viscosity (subscripts $a$ and $w$ represents air and water, respectively), $p$ is the pressure, $\boldsymbol{g}$ is the gravity, $\sigma$ is the surface tension coefficient, $\kappa$ is the curvature of the air--water interface, $\delta_s$ is a surface Dirac function that is non-zero on the interface and zero everywhere else, and $\boldsymbol{n}$ is the unit vector normal to the interface. For spatial discretization, Basilisk employs an octree-structured grid, in which each cubic cell is divided into eight smaller cells at each refinement level. The current study also utilizes the adaptive mesh refinement (AMR) function in Basilisk~\citep{van2018towards}, which dynamically refines or coarsens the grid based on the discretization error in spatial fields, i.e., $f$ and $\boldsymbol{u}$. This error is compared with a predefined threshold to determine whether a cell should be refined, coarsened, or left unchanged. More details of the numerical scheme used in Basilisk can be found in~\citet{popinet2018numerical}.

\subsection{Simulation setup}
\begin{figure*}
\centering
\includegraphics[width=\textwidth]{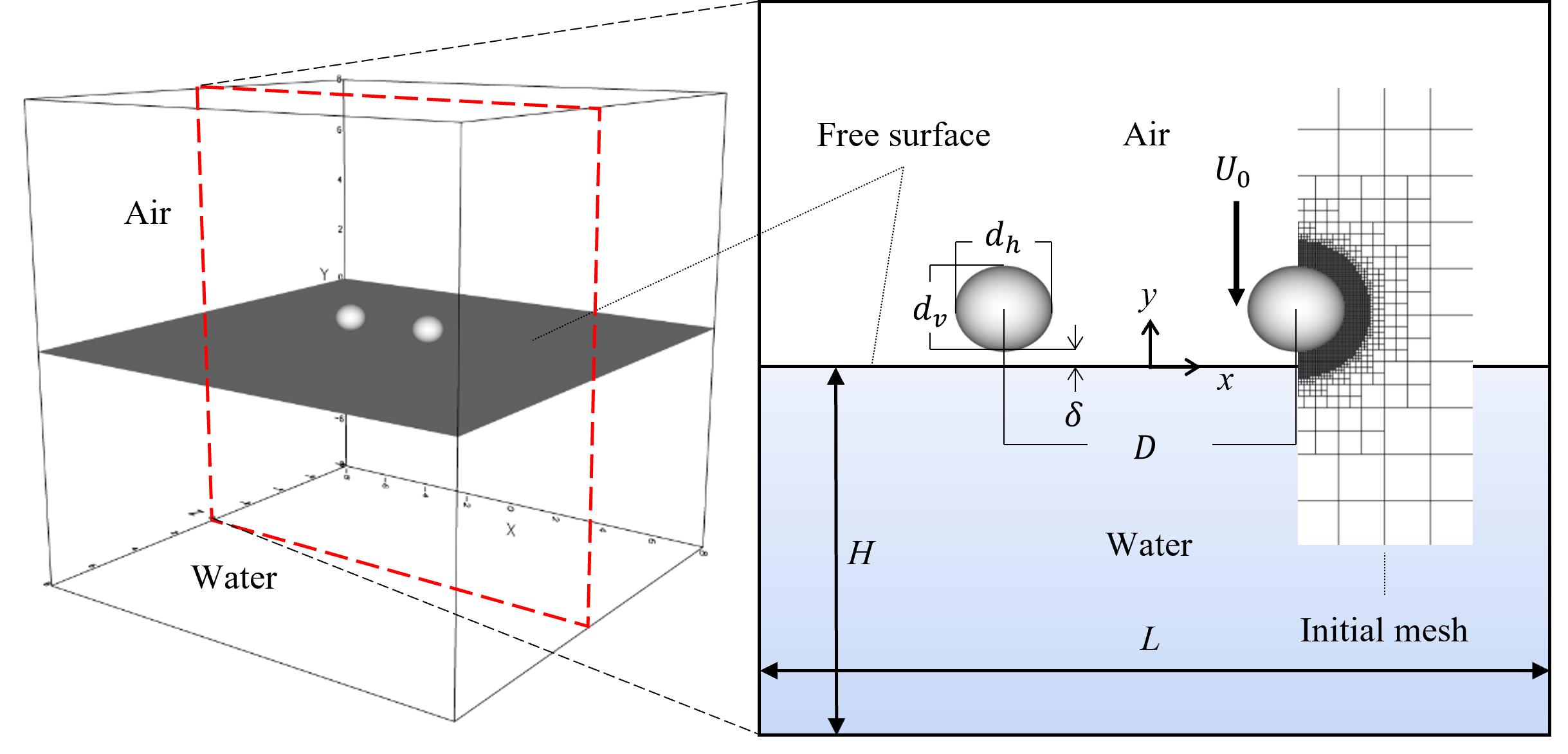}
\caption{\label{figlayout}Layout of the computational domain with a cross-sectional view in the $x$-$y$ plane illustrating the initial configuration of the raindrops and free surface in a two-raindrop case.}
\end{figure*}

In the present study, a series of raindrop impact simulations with different configurations is performed. The single-raindrop simulations with different surface tension values and diameters are labeled as SR, SR-$0.5\sigma_0$, SR-$0.25\sigma_0$, SR-$2\sigma_0$, SR-$0.25d_0$, and SR-$0.5d_0$, with $\sigma_0$ and $d_0$ denoting the surface tension and the effective raindrop diameter of case SR, respectively. Case SR serves as the reference case for comparison with all other simulations, while the remaining single-raindrop cases are performed to investigate the effect of surface tension and raindrop size on early secondary droplet formation. In addition, three two-raindrop simulations are performed with different center-to-center distances in the $x$-direction: $D=2d_h, 3d_h$, and $4d_h$, denoted as D2, D3, and D4, respectively. They are designed to investigate how raindrop interaction influences impact morphology and secondary droplet dynamics. Because the single-raindrop cases (except for SR) focus on early secondary droplet breakup, they are only simulated until 5.7 ms in physical time, whereas in cases SR, D2, and D3, the simulations are run to 45.6 ms, and D4 is extended to 47.6 ms to fully capture the interface evolution. Figure~\ref{figlayout} shows a schematic of the layout of the two-raindrop simulation domain. More details on the parameters used in each simulation case can be found in table~\ref{tab1}. 

The physical parameters of the simulations are similar to those in the laboratory experiment~\citep{murphy2015splash} and the numerical study~\citep{wang2023analysis}. The impacting raindrops are ellipsoids, with a major axis length of $d_h=4.3\text{ mm}$ and a minor axis length of $d_v=3.8\text{ mm}$. The effective diameters are defined as $d_0=(d_vd_h^2)^{1/3}=4.1\text{ mm}$. Raindrops of this size fall within the less-studied BC regime, which characterizes the most energetic air–water impact dynamics, and occur  during intense rainfall~\citep{best1950size}. The initial raindrop velocity before impact is set to $U_0 =7.2 \text{ m/s}$, which is about 82\% of the terminal velocity of raindrops this size falling in quiescent air \textcolor{black}{based on the field observations from~\citet{gunn1949terminal}, which show that the velocity of the raindrops varies in the range of 0.2 and 9 m/s and their diameters typically range from approximately 0.1 mm to 6 mm.} The initial gap between the raindrop and the free surface is set to $\delta=0.1d_0$. 

As illustrated in figure~\ref{figlayout}, the computational domain length is selected as $L=16d_0$ to ensure a balance between computational efficiency and minimization of boundary effects. The pool depth is set to $H=8d_0$, providing sufficient space above the free surface to form secondary droplets from splashing. An outflow boundary condition is applied at the top of the domain, while all other boundaries use a symmetry boundary condition. The density and viscosity ratios between water and air are $\rho_w/\rho_a=783$ and $\mu_w/\mu_a=56$, respectively. The following dimensionless numbers can be obtained, Reynolds number $Re = \rho_{w} U_0 d_0/\mu_{w} = 30069$, Weber number $We = \rho_w U_0^2 d_0/\sigma_0 = 2964$, and Froude number $Fr = U_0^2/(g d_0) = 1290$. The computational grid is initialized based on the distance to the droplet centers: the mesh is refined up to level 12 (corresponding to a smallest mesh size of $\Delta = 15.6\,\mu\text{m}$) within the raindrops and progressively coarsened with increasing distance. After initialization, the AMR algorithm adaptively refines the grid, with a maximum refinement level of $L_{max}=12$. As an illustration of the computational cost of our simulations, case D2 consumes approximately $7\times10^4$ core-hours to simulate 45.6 ms of the impact process.

\begin{table}
\begin{center}
\begin{tabular}{lc>{\color{black}}ccccccc}
Case & $N_{\mathrm{rain}}$ & $L_{\max}$ & $U_0$ (m/s) & $d_h$ (mm) & $d_v$ (mm) 
& $\sigma / \sigma_0$ & $D/d_h$ & Simulation duration (ms) \\ \hline
SR                  & 1 & 12 & 7.2 & 4.3  & 3.8  & 1    & - & 45.6\\
\textcolor{black}{SR-L10} 
& \textcolor{black}{1} 
& \textcolor{black}{10} 
& \textcolor{black}{7.2} 
& \textcolor{black}{4.3}  
& \textcolor{black}{3.8}  
& \textcolor{black}{1}    
& \textcolor{black}{-} 
& \textcolor{black}{45.6}\\
\textcolor{black}{SR-L11} 
& \textcolor{black}{1} 
& \textcolor{black}{11} 
& \textcolor{black}{7.2} 
& \textcolor{black}{4.3}  
& \textcolor{black}{3.8}  
& \textcolor{black}{1}    
& \textcolor{black}{-} 
& \textcolor{black}{45.6}\\
SR-$0.25\sigma_0$   & 1 & 12 & 7.2 & 4.3  & 3.8  & 0.25 & - & 5.7\\
SR-$0.5\sigma_0$    & 1 & 12 & 7.2 & 4.3  & 3.8  & 0.5  & - & 5.7\\
SR-$2\sigma_0$      & 1 & 12 & 7.2 & 4.3  & 3.8  & 2    & - & 5.7\\
SR-$0.25d_0$         & 1 & 12 & 7.2 & 1.08 & 0.95 & 1    & - & 5.7\\
SR-$0.5d_0$          & 1 & 12 & 7.2 & 2.15 & 1.9  & 1    & - & 5.7\\
D2                  & 2 & 12 & 7.2 & 4.3  & 3.8  & 1    & 2 & 45.6\\
D3                  & 2 & 12 & 7.2 & 4.3  & 3.8  & 1    & 3 & 45.6\\
D4                  & 2 & 12 & 7.2 & 4.3  & 3.8  & 1    & 4 & 47.6\\
\end{tabular}
\caption{\label{tab1}
Physical parameters used in the simulations. Here, $N_{\mathrm{rain}}$ is the number of raindrops, $U_0$ is the initial raindrop velocity, $d_h$ and $d_v$ are the major and minor axes of the ellipsoidal raindrop, $\sigma/\sigma_0$ is the surface tension normalized by the reference value $\sigma_0$ of case SR, and $D$ is the inter-raindrop distance.
}
\end{center}
\end{table}

\subsection{Validation}

\begin{figure}
\centering
\includegraphics[width=0.9\textwidth]{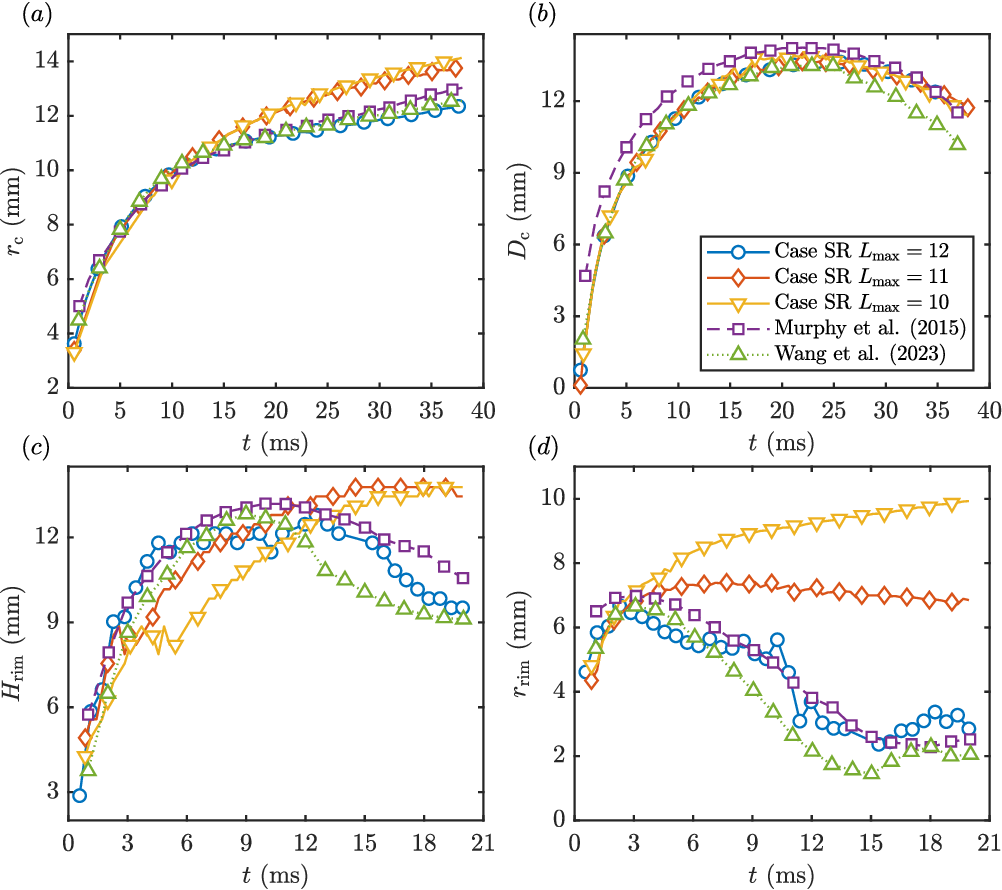}
\caption{\label{figval}
\textcolor{black}{Time evolution of the (a) cavity radius, (b) cavity depth, (c) crown rim height, and (d) crown rim radius in case SR for $L_{\max}=10$, 11, and 12. Also plotted are reference results from~\citet{murphy2015splash} and~\citet{wang2023analysis}}}
\end{figure}

\textcolor{black}{For validation, simulation data from the single raindrop case SR are used to calculate four quantities representative of the impact morphology: the cavity radius $r_c$, defined as the radius of the circular interface at $y=0$, the cavity depth $D_c$, defined as the distance between the deepest point of the cavity interface and the free surface at $y=0$, the crown rim height $H_{\mathrm{rim}}$, defined as the distance between the base of the ligaments and the free surface at $y=0$, and the crown rim radius $r_{\mathrm{rim}}$, defined as the radial distance from the impact center to the crown rim at the identified rim height. The full procedure for extracting $r_c$, $D_c$, $H_{\mathrm{rim}}$, and $r_{\mathrm{rim}}$ is detailed in appendix~\ref{app:impact_geometry}. To assess grid convergence and grid independence, the time histories of $r_c$, $D_c$, $H_{\mathrm{rim}}$, and $r_{\mathrm{rim}}$ for the SR case are compared using $L_{\max}=10$, 11, and 12, as shown in figure~\ref{figval}. While the results for $L_{\max}=10$ and 11 deviate from those reported by~\citet{murphy2015splash} and~\citet{wang2023analysis}, the results for $L_{\max}=12$ show good agreement with these reference studies, suggesting that $L_{\max}=12$ is sufficient for capturing the impact morphology in the present study.} 

\begin{figure}
\centering
\includegraphics[width=0.55\textwidth]{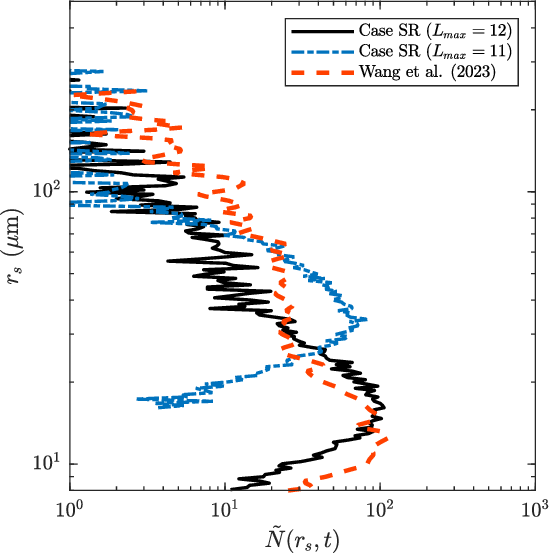}
\caption{
\textcolor{black}{Time-averaged secondary droplet size distributions extracted from case SR between 0.57--4 ms for $L_{\max}=11$ and $L_{\max}=12$, compared with the results of~\citet{wang2023analysis} averaged between 0.2--4 ms.}
}
\label{figvalsec}
\end{figure}

\textcolor{black}{In figure~\ref{figvalsec}, we compare the secondary droplet size distribution averaged over $0.57$--$4~\mathrm{ms}$ with those obtained over $0.2$--$4~\mathrm{ms}$ by~\citet{wang2023analysis} who used a higher resolution. A notable improvement in our distribution is observed when $L_{\max}$ is increased from $11$ to $12$. The remaining difference between our result and theirs is primarily attributed to temporal variability in the droplet production: for instance, their results have shown that the droplet number can vary by up to one order of magnitude between two time windows $0.2$--$2~\mathrm{ms}$ and $3$--$4~\mathrm{ms}$. Overall, this comparison suggests that the resolution used in the present study is sufficient to capture the dominant secondary droplet population.}

\section{Results}
\label{sec:results}
\subsection{Early generation of secondary droplets}
In this section, we examine the early generation of secondary droplets from raindrop impacts. The focus is on the early stage of raindrop--pool interaction within $5.7\ \mathrm{ms}$ after release, when most secondary droplets are produced. Following the discussion of mesh resolution on atomization and droplet generation from~\citet{pairetti2020mesh}, the droplet statistics in the present study only include droplets with diameters larger than $64\,\mu\text{m}$, approximately four times the size of the smallest cell. Secondary droplets smaller than this value are considered not well resolved for our mesh setup and are not discussed here. 

\subsubsection{Possible mechanism for secondary droplet production}

Historically, the Rayleigh--Plateau instability~\citep{rayleigh1892xvi,bremond2006atomization}, the Richtmyer--Meshkov instability~\citep{richtmyer1960taylor,gueyffier1998finger}, and a combined Richtmyer--Meshkov and Rayleigh--Taylor mechanism~\citep{rayleigh1882investigation,krechetnikov2009crown} have been proposed as possible candidates for governing droplet generation from the crown rim. \textcolor{black}{Since previous studies have attributed secondary droplet production during drop--pool impact to Rayleigh--Plateau instability of the crown rim~\citep{roisman2006spray,zhang2010wavelength,constante2023impact,Anirudh_Behera_Sahu_2026}, we first examine whether this mechanism can explain the secondary droplet generation observed here. Following~\citet{zhang2010wavelength}, we compare the measured crown rim corrugation spectrum with the Rayleigh--Plateau prediction based on the instantaneous crown radius and rim thickness, as described in appendix~\ref{app:crown_corrugation}. Our calculations indicate that crown breakup in the present high-$Re$ and high-$We$ regime is not governed by direct Rayleigh--Plateau growth on the crown rim.}

\textcolor{black}{Using the impacting raindrop diameter and velocity in the present simulations gives $Re_p=U_0d_0/\nu_a\approx 2000$, placing the airflow around the falling raindrop in the unsteady and turbulent-wake regime~\citep{mathai2020bubbly}. Thus, the early impact and crown development occur in a strongly inertial gas-flow environment, in which the breakup process is likely affected by unsteady airflow. In the fragmentation framework reviewed by~\citet{villermaux2007fragmentation}, the lognormal distribution is associated with a sequential cascade breakup mechanism~\citep{Kolmogorov1941}, whereas gamma-type distributions are linked to ligament-mediated breakup and the coalescence and rearrangement of corrugations along ligaments~\citep{bremond2006atomization}. Following these ideas, we have examined whether the normalized droplet radius distributions could be represented by lognormal or gamma-type forms. However, neither provides a robust description across the cases and times considered (not shown). In fact, the difference between these two distributions in quantifying droplets generated during fragmentation is relatively small~\citep{villermaux2007fragmentation}. Therefore, distinguishing which of these two mechanisms governs the present problem will require additional simulations to obtain droplet statistics with reduced scatter in future studies.}

\subsubsection{Scaling law derivation and validation}

\textcolor{black}{We use the Kolmogorov--Hinze framework \citep{Kolmogorov1949,Hinze1955} to guide the scaling analysis of the secondary droplet size distribution. This approach provides a dimensional argument for how droplet production may be controlled by the competition between inertial fluctuations and surface tension. In the present analysis, it is used to describe the surface-tension-controlled part of the spectrum. We do not attempt to derive a separate inertia-dominated scaling for larger droplets because the sample size is limited and the scatter is larger. } One notable application of this framework is found in studies of air bubbles in breaking waves, where different scaling laws were identified for bubbles larger~\citep{garrett2000connection} and smaller~\citep{deane2002scale} than the Hinze scale. Inspired by these seminal works, we derive a power scaling law for the size spectrum of secondary droplets $N_d$, i.e., the number of secondary droplets per unit droplet radius increment per unit volume of air. It is assumed that $N_d$ is a multiplicative function of the rate of water being ejected into the air $Q$, i.e., the volume of water ejected per volume of air per second, surface tension $\sigma$, water density $\rho_w$, and droplet radius $r_s$. After equalizing the dimensions of these parameters, the following relationship between $N_d(r_s)$ and the droplet radius $r_s$ can be obtained:
\begin{equation}
N_d(r_s)\propto Q(\sigma/\rho_w)^{-1/2}r_s^{-5/2}.
\label{eq:spectrum}
\end{equation}

\textcolor{black}{It is worth noting that, while~\eqref{eq:spectrum} is similar to that derived by~\citet{deane2002scale}, we do not introduce the raindrop impact velocity as an additional independent parameter in the present scaling. Here, $Q$ is treated as the impact-driven water ejection rate and can therefore vary with the impact conditions, including the impact velocity. Thus, the present scaling should not be interpreted as assuming that velocity effects are negligible; rather, the effect of velocity is incorporated through its influence on $Q$.}

\begin{figure}
\centering
\includegraphics{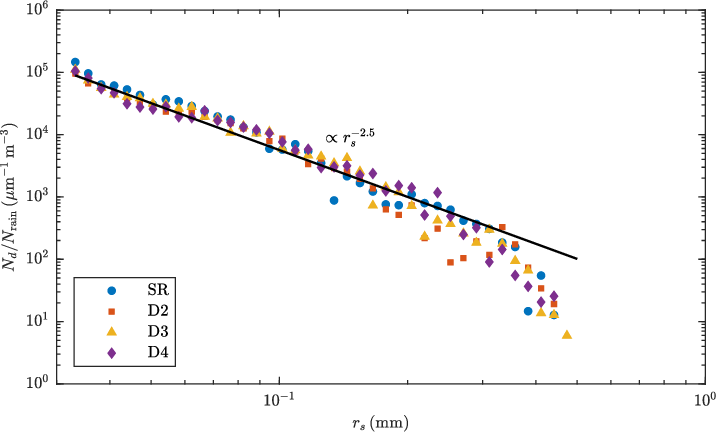}
\caption{\label{figND_singledouble}Time-averaged number density spectra as a function of the droplet radius, normalized by the number of raindrops $N_{\mathrm{rain}}$ for cases SR, D2, D3, and D4.}
\end{figure}

To validate~\eqref{eq:spectrum}, the time-averaged secondary droplet number density spectra are first computed for cases SR, D2, D3, and D4 over the interval $t = 0.57$ ms to $t = 5.7$ ms. Figure~\ref{figND_singledouble} shows the result normalized by the number of raindrops $N_{\mathrm{rain}}$. The collapse of the data among different cases confirms a negligible effect associated with the raindrop interaction. More importantly, all four cases follow the predicted $N_d(r_s) \propto r_s^{-5/2}$ scaling well.

\begin{figure}
\centering
\includegraphics[width=\textwidth]{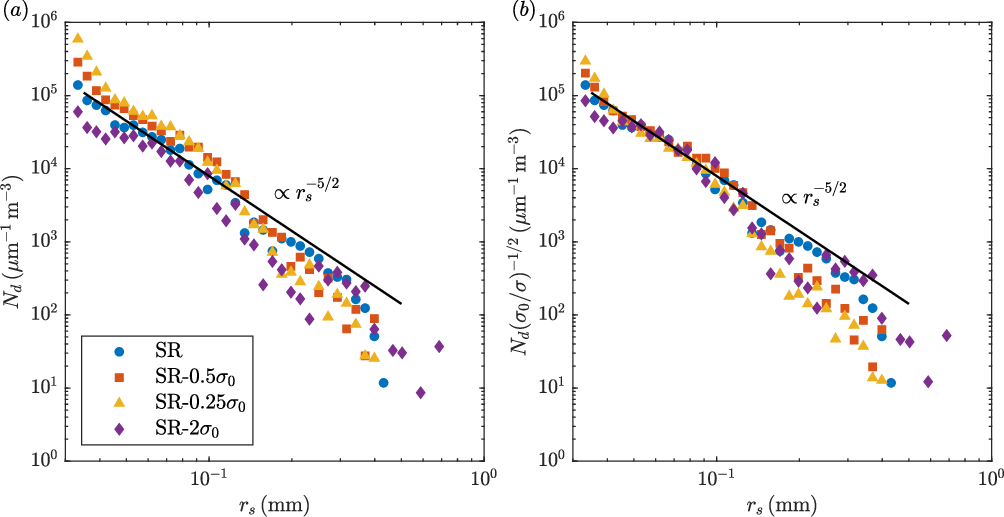}
\caption{\label{figND_sigma}(a) Time-averaged number density spectra of secondary droplets and (b) the corresponding rescaled spectra as functions of droplet radius for single-raindrop cases with different surface tensions.}
\end{figure}

\begin{figure}
\centering
\includegraphics[width=\textwidth]{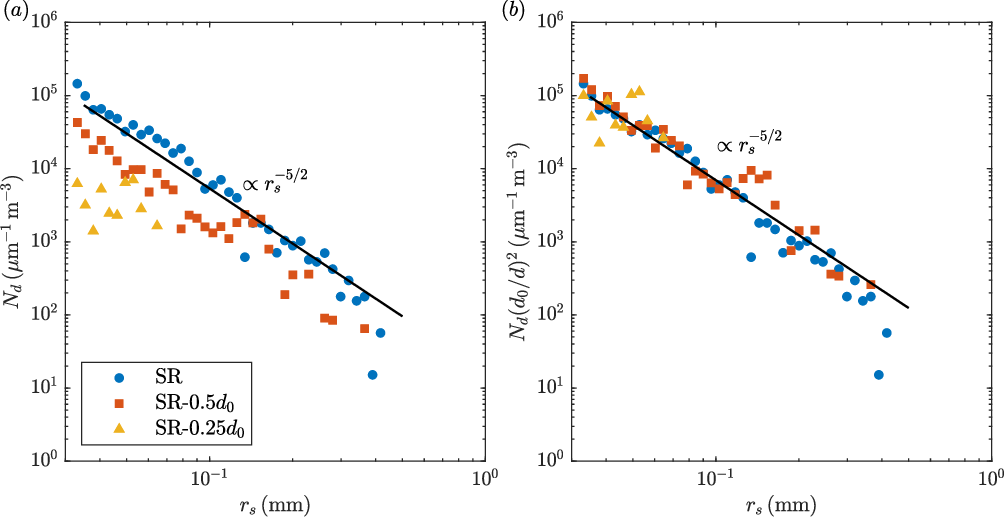}
\caption{\label{figND_d}Same as figure~\ref{figND_sigma} but for single-raindrop cases with different raindrop diameters.}
\end{figure}

A similar strategy is adopted for investigating the impact of surface tension on the secondary droplet distribution. Their time-averaged spectra are shown in figure~\ref{figND_sigma}(a), demonstrating the impact caused by varying the surface tension. Figure~\ref{figND_sigma}(b) shows the rescaled spectra $N_d(r_s)(\sigma_0/\sigma)^{-1/2}$, which collapse well in the range of $0.05\space \mathrm{mm}<r_s <0.1\space \mathrm{mm}$ for the cases with different surface tension, thus confirming the scaling, i.e., $N_d(r_s) \propto \sigma^{-1/2}$. Finally, we examine how the raindrop size affects the secondary droplet distribution. Figure~\ref{figND_d}(a) shows the time-averaged spectra of the single-raindrop simulations with varying raindrop diameters, including cases SR, SR-$0.5d_0$, and SR-$0.25d_0$ (see table~\ref{tab1}). All cases follow the $N_d(r_s) \propto r_s^{-5/2}$ scaling closely, indicating that the underlying breakup mechanism remains similar across different raindrop sizes.

\begin{figure}
\centering
\includegraphics[width=0.5\textwidth]{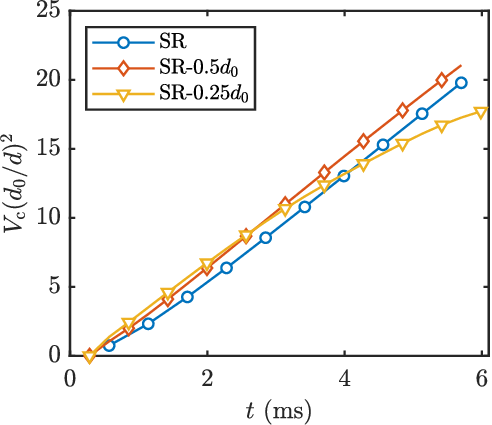}
\caption{\label{fig_crown_volume} \textcolor{black}{Time history of the nondimensionalized crown volume rescaled by $(d_0/d)^2$ for single-raindrop cases with different drop diameters.}}
\end{figure}

\textcolor{black}{For different raindrop diameters, the secondary droplet number density continues to follow the same $r_s^{-5/2}$ scaling. The main effect of increasing $d$ is instead an increase in the magnitude of $N_d$. As shown in figure~\ref{figND_d}(b), the spectra collapse well when the number densities are multiplied by $(d_0/d)^2$, suggesting a dependence of $N_d\propto d^2$. This scaling can be interpreted from the crown volume: for a thin crown sheet, the volume is expected to scale approximately with the raindrop impact footprint, which is proportional to $d^2$. The resulting increase in crown volume provides more liquid available for breakup into secondary droplets, consistent with the collapse of the crown volume time histories in figure~\ref{fig_crown_volume}.}

We briefly discuss the scatter observed in the simulation data and its possible causes. For example, the smallest secondary droplets approach the under-resolved size limit of the mesh, which can lead to over-counting, as noted by~\citet{pairetti2020mesh}. As expected, this numerical issue becomes more significant as the surface tension decreases (comparing $N_d$ for $r_s < 0.04\space \mathrm{mm}$ in figure~\ref{figND_sigma}a and~\ref{figND_sigma}b). On the other hand, the scatter also tends to increase for droplets larger than approximately $0.1$ mm because fewer larger droplets are generated. For instance, there are approximately \textit{O}(10) secondary droplets with a radius of $0.2$ mm in the computational domain. Because of the scatter, we do not attempt to derive a scaling law for droplets larger than the Hinze scale, for which breakup induced by inertial fluctuations is expected to dominate over surface tension. 

\subsection{Impact morphology}
\begin{figure}
\centering
\includegraphics[width=\textwidth]{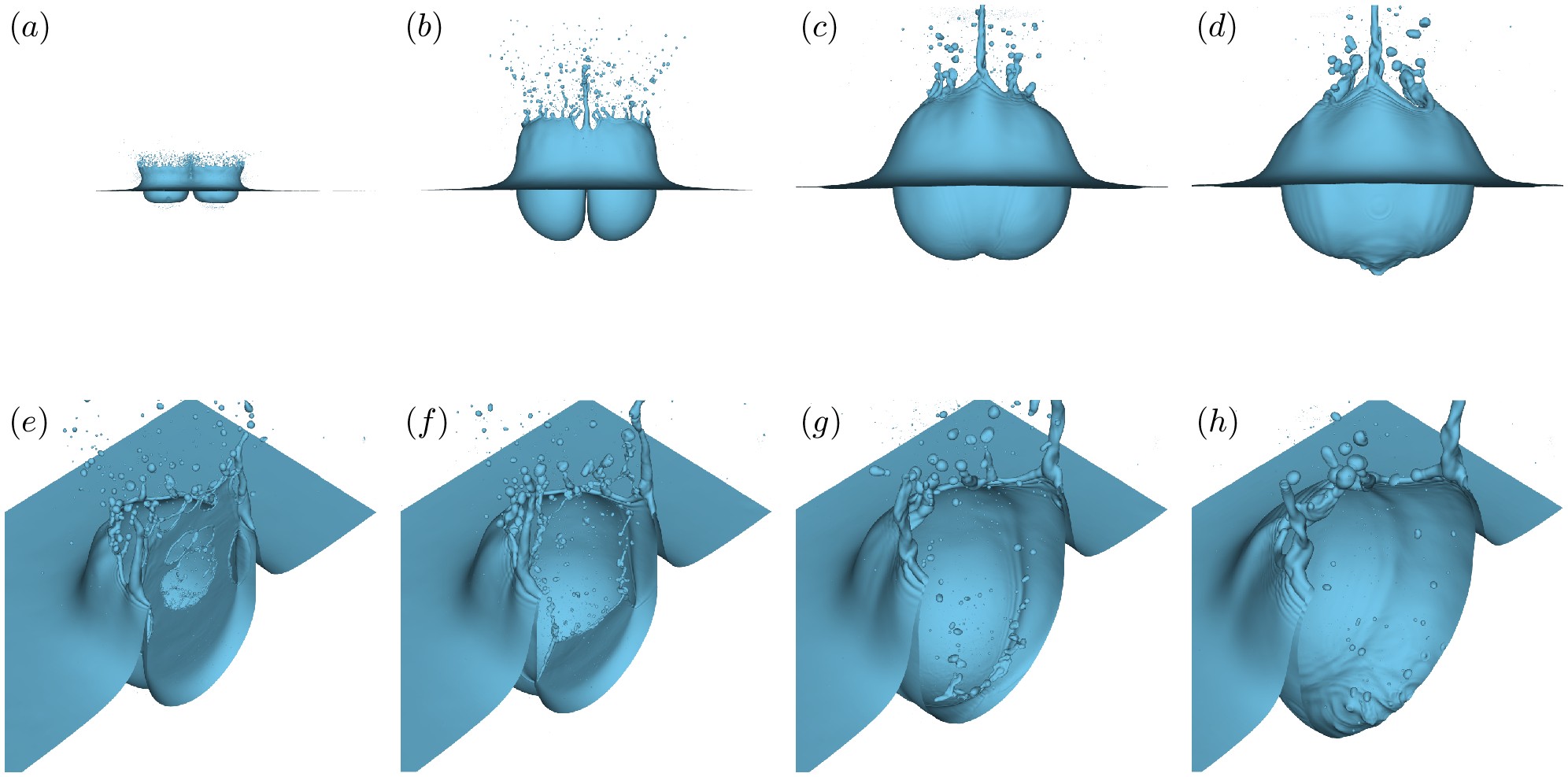}
\caption{\label{fig2d}(a-d) Side views and (e-h) isometric views of the impact morphology in case D2. In (e-h), only part of the domain is shown for clarity. The time instants in (a–d) are 1.14 ms, 6.27 ms, 17.67 ms, and 27.65 ms, respectively, whereas those in (e–h) are 8.83 ms, 11.68 ms, 17.67 ms, and 27.65 ms.}
\end{figure}

\begin{figure}
\centering
\includegraphics[width=\textwidth]{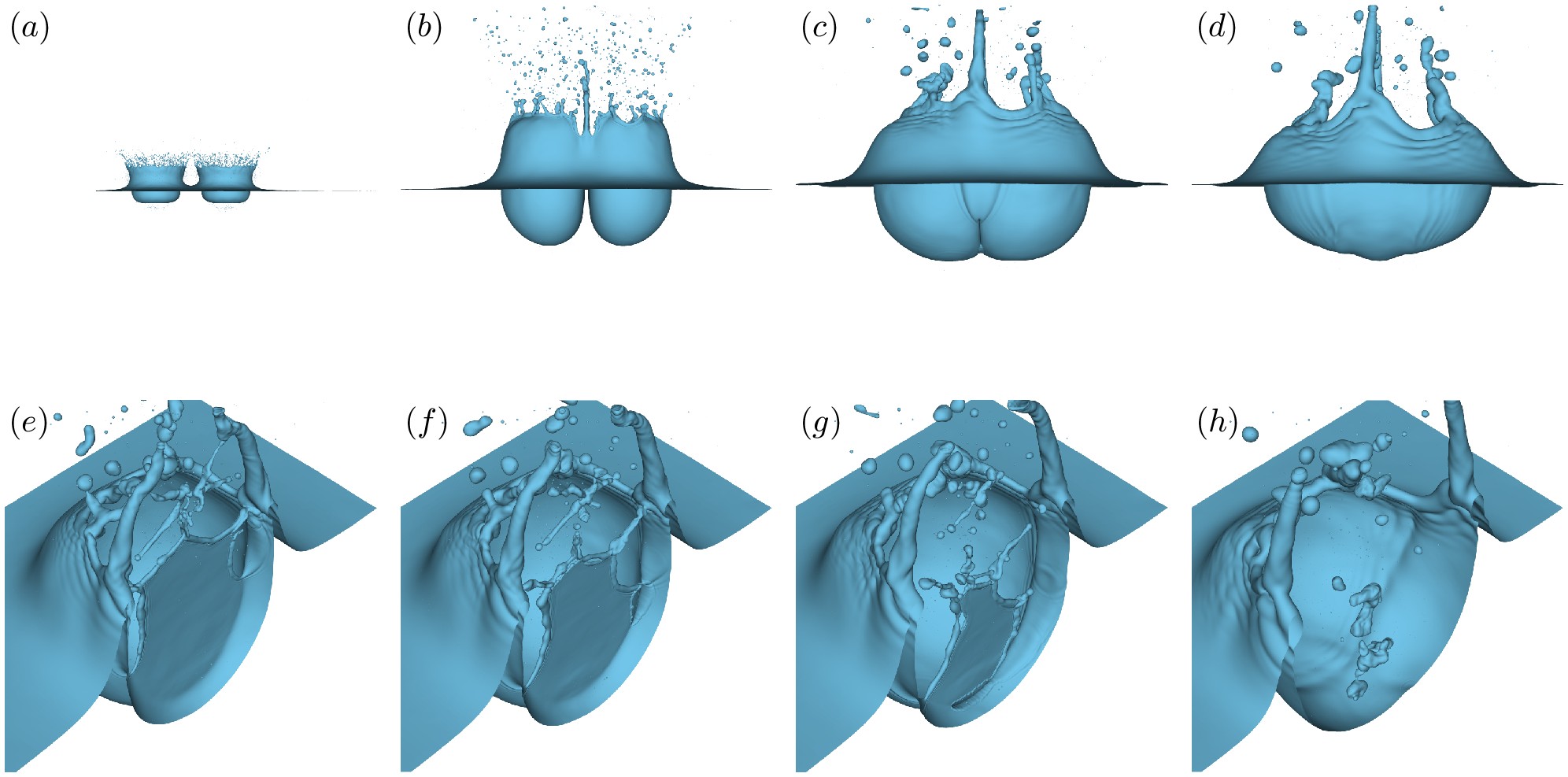}
\caption{\label{fig3d}Same as figure~\ref{fig2d}, but for case D3. The time instants in (a–d) are 1.14 ms, 7.98 ms, 29.93 ms, and 41.33 ms, respectively, whereas those in (e–h) are 25.08 ms, 27.36 ms, 29.93 ms, and 41.33 ms.}
\end{figure}

\begin{figure}
\centering
\includegraphics[width=\textwidth]{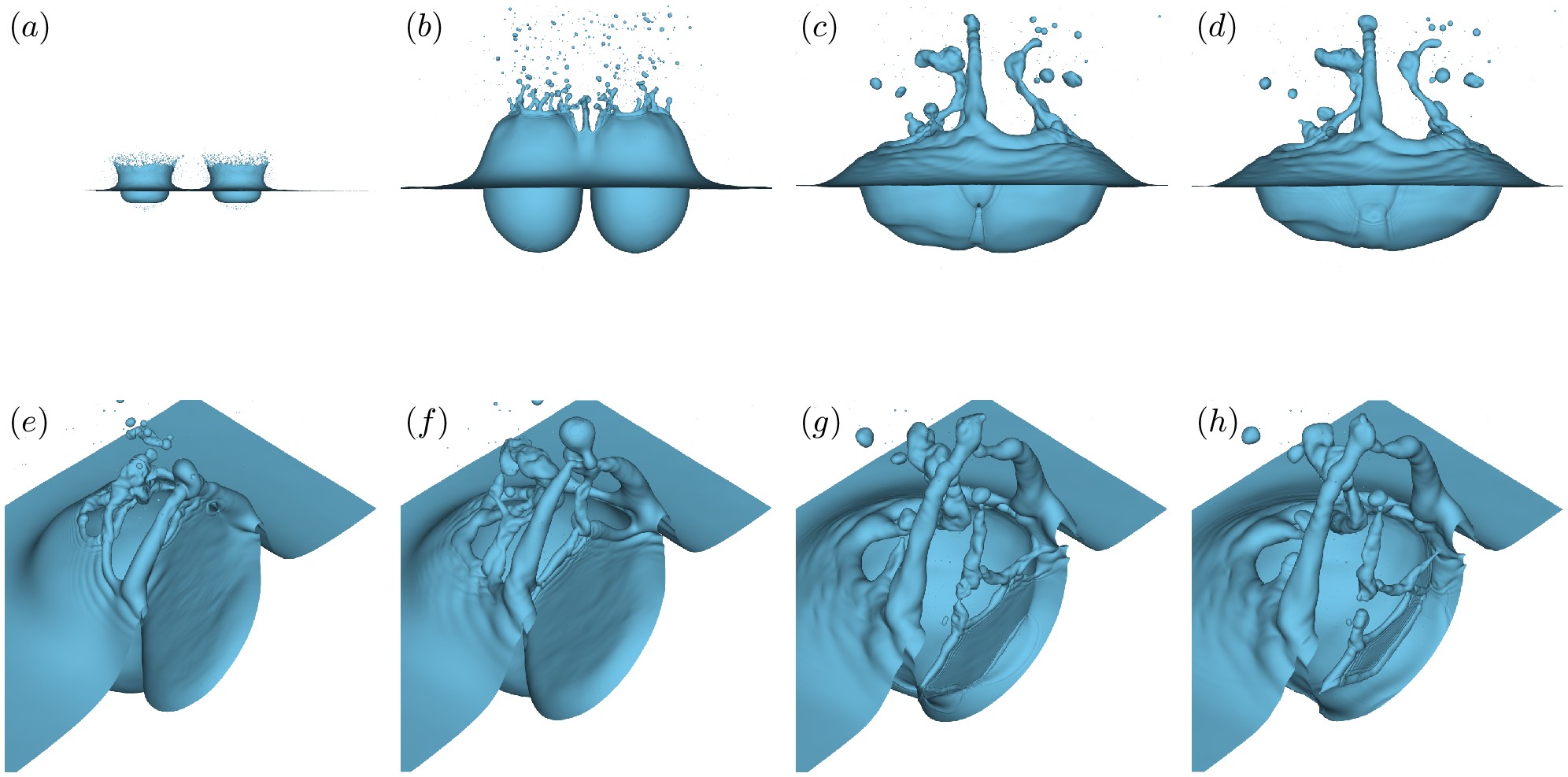}
\caption{\label{fig4d}Same as figure~\ref{fig2d}, but for case D4. The time instants in (a–d) are 1.14 ms, 11.4 ms, 45.03 ms, and 47.6 ms, respectively, whereas those in (e–h) are 22.8 ms, 33.92 ms, 45.03 ms, and 47.6 ms.}
\end{figure}

In this section, the morphology of the impact is analyzed. As the single-raindrop case has been extensively studied, the focus here is on the two-raindrop cases. For all cases, $t = 0$ is defined as the start of the simulation, when the raindrops are initially positioned above the free surface (figure~\ref{figlayout}). Figure~\ref{fig2d}, \ref{fig3d}, and \ref{fig4d} illustrate several stages of the impact process for case D2, D3, and D4. First, upon impact, the raindrops create flat-bottomed cavities resembling those seen in single-raindrop impacts. These cavities gradually transition into hemispherical shapes while producing cylindrical upward liquid sheets, or crowns~\citep{thoroddsen2002ejecta, deegan2007complexities, zhang2012evolution}. Second, the crowns merge, forming a central liquid film, a feature observed in previous multi-drop impact studies~\citep{li2016simulation, liang2018simultaneous, fest2021multiple, poureslami2023simultaneous, zhou2024numerical}. Third, the subsurface cavities interact and eventually merge into a single large cavity after the central film fully ruptures and detaches from the cavity wall. Finally, the release of surface tension following the detachment of the central film from the cavity bottom leads to a recoil that generates a localized downward bulge at the cavity bottom, further deepening the cavity. Notably, in high-speed single-raindrop impacts, it is common for the crown to converge radially inward and pinch off, creating a bubble canopy~\citep{worthington1883impact, engel1966crater, bisighini2010crater, murphy2015splash}. In the two-raindrop cases (D2, D3, and D4), although the crowns also curve inward and partially merge, they do not fully enclose at the center. As a result, the characteristic jetting behavior observed after bubble canopy formation in single-raindrop cases is absent here.

Comparing the side views shown in figure~\ref{fig2d}, \ref{fig3d}, and \ref{fig4d}, it is clear that increasing the inter‑raindrop distance progressively delays the merging of crowns and cavities, yielding a weaker downward bulge and shallower penetration depth. This trend is closely linked to the central film behavior. With larger spacing, the film disintegration (depicted on the bottom rows of figure~\ref{fig2d}, \ref{fig3d}, and \ref{fig4d}) occurs later and with less intensity, which reduces the recoil upon detachment from the cavity base and suppresses bulge formation.

To examine the temporal evolution of cavity depth, we first outline the analytical solution for a single-raindrop impact formulated by~\citet{bisighini2010crater}. In this model, based on potential flow theory, the velocity potential of the downward-expanding cavity is treated as a superposition of flow past a sphere and a radially expanding flow. The pressure distribution at the cavity surface is determined using the Bernoulli equation, and a dynamic boundary condition is applied to account for capillary forces and gravity. These lead to a system of ordinary differential equations governing the evolution of the cavity geometry: 
\begin{eqnarray}
\ddot{\alpha} & = & -\frac{3 \dot\alpha^2}{2 \alpha} - \frac{2}{\alpha^2 We} - \frac{1}{Fr}\frac{\zeta}{\alpha} + \frac{7 \dot\zeta^2}{4 \alpha},  \label{eq:bisighini1} \\[10pt]
\ddot{\zeta} & = & -3\frac{ \dot\alpha \dot{\zeta}}{\alpha} - \frac{9}{2}\frac{ \dot\zeta^2}{ \alpha} - \frac{2}{Fr}. \label{eq:bisighini2}
\end{eqnarray}
where $\alpha$ and $\zeta$ are the cavity radius and the axial position of the cavity center, normalized by the initial droplet diameter. The initial conditions for this model are derived from the pressure distribution at the impact axis and the free surface of the cavity, resulting in:  
\begin{equation}
\dot{\alpha} \approx 0.17, \quad \alpha \approx \alpha_0 + 0.17 \tau, \quad 
\dot{\zeta} \approx 0.27, \quad \zeta \approx -\alpha_0 + 0.27 \tau, 
\label{eq:initial_conditions}
\end{equation}
where $\tau=U_0t/d_0$ is the non-dimensional time and $\alpha_0$ is a constant associated with the cavity radius $r_c$. To determine $\alpha_0$, one can use its connection to the dimensionless cavity diameter $\Omega=2r_c/d_0$:  
\begin{equation}
\Omega = 2 \sqrt{\alpha^2 - \zeta^2} \approx 2 \sqrt{(\alpha_0 + 0.17 \tau)^2 - (0.27 \tau - \alpha_0)^2}. 
\label{eq:omega}
\end{equation}
By fitting the cavity radius data gathered from our SR simulation (refer to figure~\ref{figval}a), $\alpha_0\approx 0.81$ is obtained. Note that in the derivation by~\cite{bisighini2010crater}, the initial condition is determined to be at $\tau=2$ because it marks the transition from complex interaction between the droplet and pool surface to single-fluid inertial flow, which is more suitable for potential flow analysis. Therefore, the initial conditions for~\eqref{eq:bisighini1} and~\eqref{eq:bisighini2} are: \( \alpha(2) = 1.15 \), \( \dot{\alpha}(2) = 0.17 \), \( \zeta(2) = -0.27 \), and \( \dot{\zeta}(2) = 0.27 \). Solving~\eqref{eq:bisighini1} and~\eqref{eq:bisighini2}, we can obtain the analytical cavity depth for a single-raindrop impact $\Delta=\alpha+\zeta$.

\begin{figure}
\centering
\includegraphics{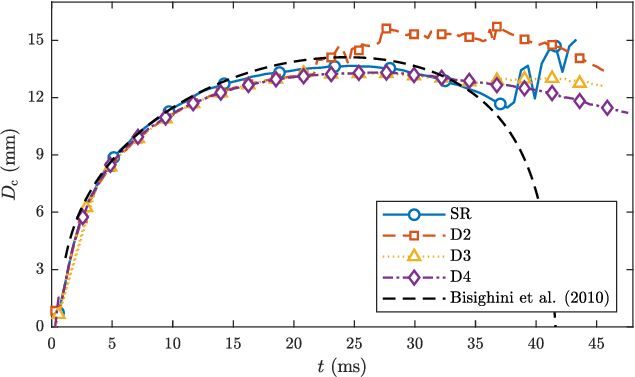}
\caption{\label{figdepth}Comparison of cavity depth evolution for single and two-raindrop impact cases against the analytical model. }
\end{figure}

Figure~\ref{figdepth} compares the temporal evolution of cavity depth in cases SR, D2, D3, and D4. The cavity depths for all cases initially follow similar trends and agree well with the analytical prediction up to approximately 20~ms, beyond which notable differences emerge depending on the configuration. In case D2, cavity depth increases significantly after 21.7~ms and continues to grow until 27.6~ms, after which the growth stagnates. Following 36.8~ms, the cavity begins to retract, leading to a corresponding decrease in depth. Case D3 reaches its first peak cavity depth at 27.6~ms, followed by a slight decrease due to cavity stagnation and retraction. Upon bulge formation, there is a modest increase in depth after 35.9~ms, reaching a second peak at 41.3~ms before retracting again. In its first retraction phase between 27.6~ms and 35.9~ms, the cavity in case D3 retracts more slowly than that in case SR. This is attributed to the proximity and subsequent merging of its cavities. The cavity depth of the D4 case remains nearly identical to the D3 case up to 35.9~ms. Since bulge formation is weak, no second peak is observed. Note that the SR case shows an irregular increase in cavity depth after 36~ms, which results from the downward liquid jet that penetrates the cavity bottom captured by our cavity depth algorithm (appendix~\ref{app:impact_geometry}).

\begin{figure}
\centering
\includegraphics{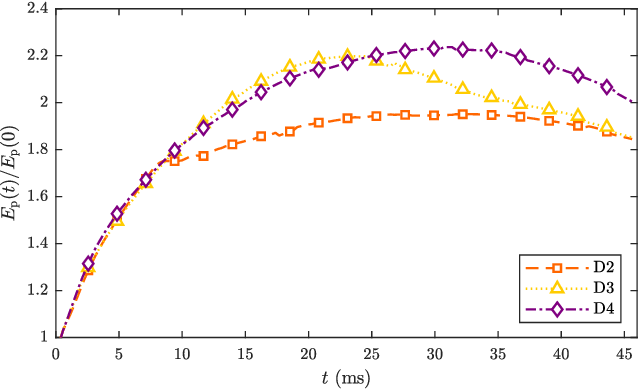}
\caption{\label{figSE}Comparison of the normalized pool surface energy $E_p(t)/E_p(0)$ with time between D2, D3, and D4.}
\end{figure}

In addition, to quantify the evolution of the morphology, we calculate the pool surface energy, $E_p$, defined as the surface area of the largest connected fluid region in the domain multiplied by $\sigma_0$.  As shown in figure~\ref{figSE}, the normalized pool surface energy $E_p(t)/E_p(0)$ remains nearly identical across all cases until approximately 9~ms. Beyond this time, case D2 exhibits a slower increase compared to the others, indicating fluid detachment from the pool associated with its rapid central film disintegration (figure~\ref{fig2d}e–g). For case D3, a sharper decline appears around 25~ms, corresponding to the onset of central film breakup (figure~\ref{fig3d}e–g). In contrast, because case D4 shows no significant central film disintegration, its $E_p$ follows the expected trend of cavity expansion and retraction without any notable deviation.  

\subsection{Raindrop interaction and droplet evolution}
\textcolor{black}{In this section, the droplet evolution is investigated, with a focus on how inter-raindrop interaction in cases D2, D3, and D4 shapes secondary droplet production, spatial distribution, and re-merging dynamics. Our analysis of secondary droplet behavior extends up to 15~ms, as they are no longer produced in significant numbers beyond this time. We distinguish between the production of droplets, their redistribution above the free surface, and their transport into the cavity, since these processes are affected differently by the central liquid film and cavity air flow.}

\subsubsection{Droplet production and spatial distribution}

\begin{figure}
\centering
\includegraphics[width=0.9\textwidth]{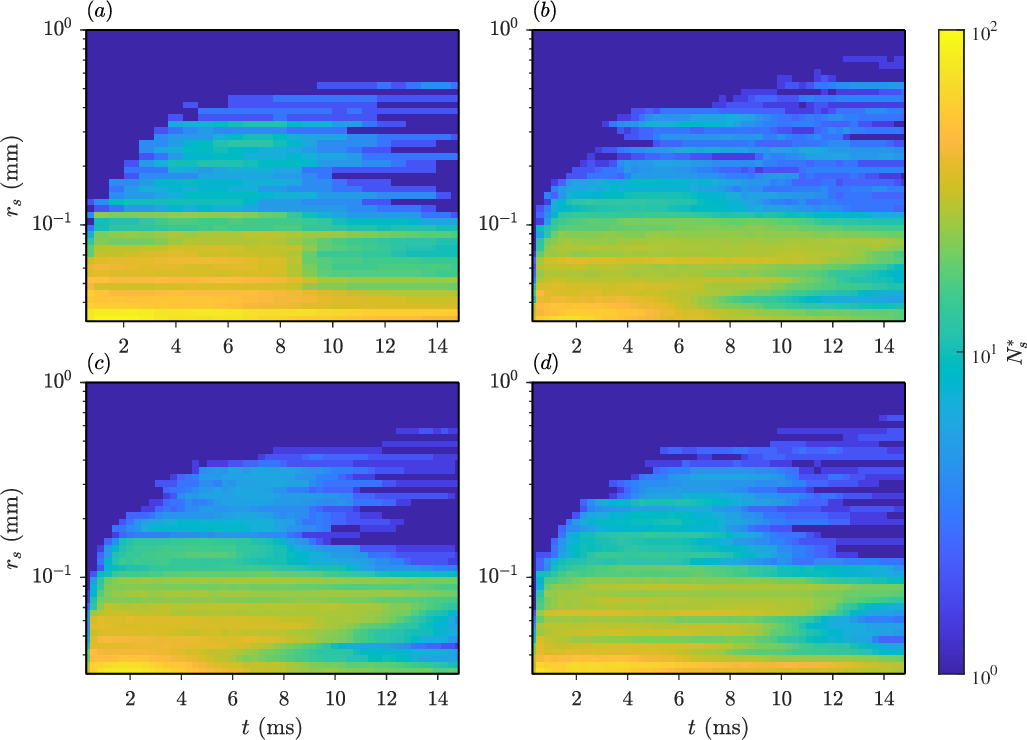}
\caption{\label{fig_tot_size}Temporal evolution of the normalized number of secondary droplets at different sizes for case (a) SR, (b) D2, (c) D3, and (d) D4.}
\end{figure}

\textcolor{black}{The temporal evolution of $N_s^*=N_s/N_{\mathrm{rain}}$, the number of secondary droplets normalized by the number of raindrops, at different sizes in cases SR, D2, D3, and D4 is shown in figure~\ref{fig_tot_size}. This size-time distribution provides an overview of when droplets of different sizes are produced and how long they remain in the domain. Overall, all cases demonstrate similar trends: small secondary droplets are produced early in the impact in large quantities, while larger droplets appear later in smaller quantities. This indicates that the general sequence of droplet production is similar across the single- and two-raindrop cases. The main spacing-dependent difference appears at later times, where case D2 shows a higher $N_s^*$ for droplets with radii $r_s$ in the range $0.1<r_s<0.5~\mathrm{mm}$ during $9<t<15~\mathrm{ms}$. These secondary droplets originate from central film disintegration, which is absent in SR. As discussed previously, the central film rupture for cases D3 and D4 is not violent enough to create secondary droplets, leading to the lower $N_s^*$ in this size range.
}

\begin{figure}
\centering
\includegraphics[width=\textwidth]{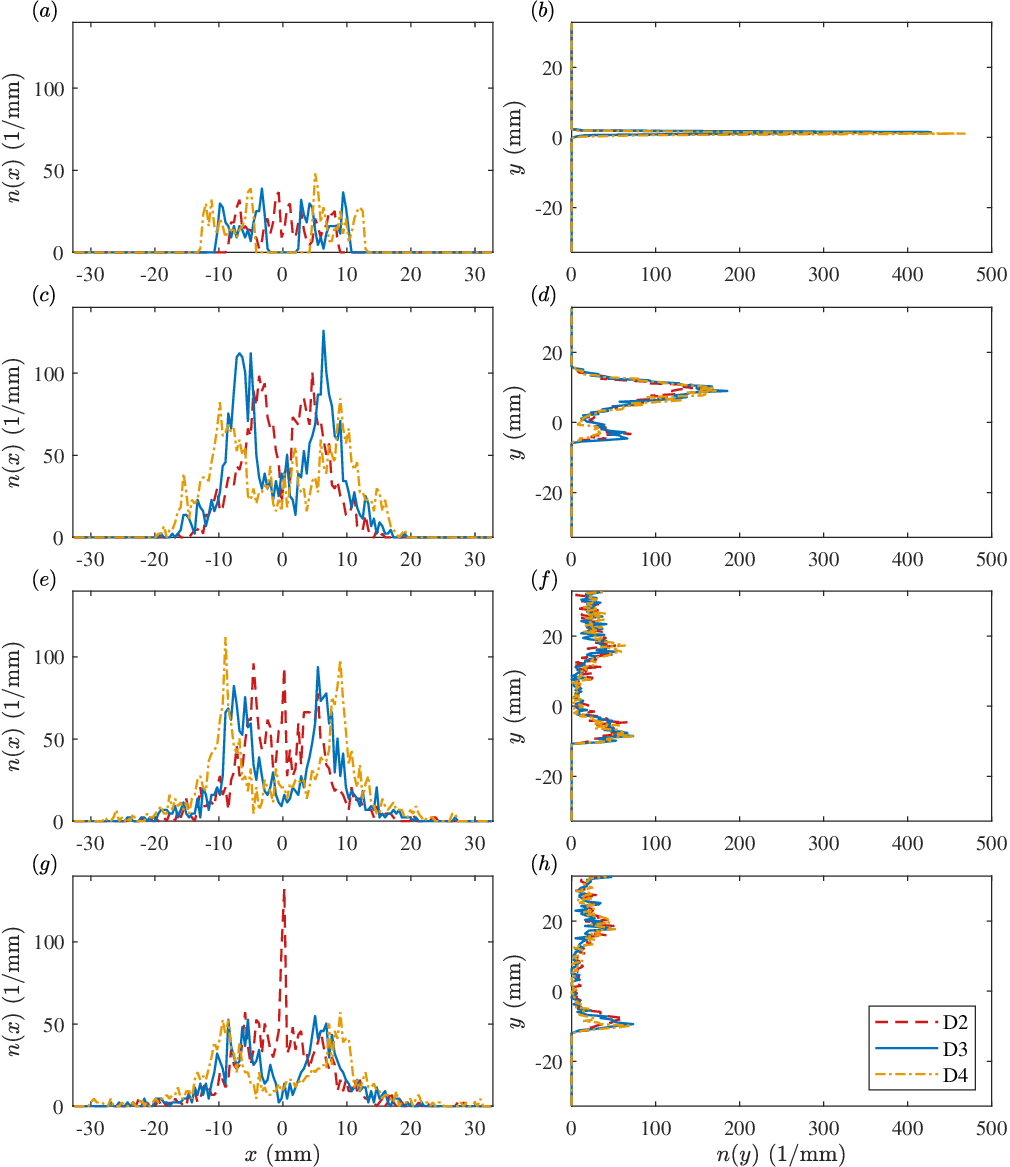}
\caption{\label{fig_xy_den} Number density profiles of secondary droplets along the x- and y-directions for D2, D3, and D4 cases at (a, b) 0.28 ms, (c, d) 2.56 ms, (e, f) 8.83 ms, and (g, h) 11.68 ms.}
\end{figure}

\textcolor{black}{To determine whether the late-time larger-droplet population in D2 is indeed associated with central film disintegration}, their spatial number density profiles along the $x$- and $y$-directions are plotted at selected times from 0.28~ms to 11.68~ms in figure~\ref{fig_xy_den}. Here, $n(x)$ and $n(y)$ denote the number density of secondary droplets in the $x$- and $y$-directions per unit distance, respectively. \textcolor{black}{The $x$-direction profile is particularly useful because droplets produced by central film disintegration are expected to appear near the mid-plane between the two impact sites.}

The behavior of $n(x)$ can be roughly divided into two stages, before and after the central film disintegration. At 0.28~ms, relatively uniform distributions, appearing as plateaus, are observed (figure~\ref{fig_xy_den}a). For cases D3 and D4, two distinct plateaus are observed, reflecting their larger initial raindrop spacing, while case D2 exhibits a single plateau due to its smaller inter-raindrop distance. A twin-peak distribution appears in all cases (figure~\ref{fig_xy_den}c) as droplets fall towards the two cavities. At 8.83~ms, although cases D3 and D4 maintain a two-peak distribution (figure~\ref{fig_xy_den}e), case D2 develops a third peak near $x=0$, which is attributed to central film disintegration producing additional droplets. By 11.68~ms (figure~\ref{fig_xy_den}g), the profiles show a prominent central peak for case D2, indicating that most of its secondary droplets in the domain at this time were produced by central film disintegration. In contrast, cases D3 and D4 still maintain a two-peak distribution, although with significantly fewer droplets overall. \textcolor{black}{Therefore, the appearance of the central peak in D2 provides a spatial signature of secondary droplet production by central film disintegration, while the two-peak distributions in D3 and D4 indicate that their droplet populations remain primarily associated with the two original impact sites.}

For $n(y)$, the initial profiles (figure~\ref{fig_xy_den}b) are sharply concentrated near the unperturbed free surface, $y=0$, where the droplets are generated. By 2.56~ms (figure~\ref{fig_xy_den}d), the peak spreads in both directions and droplets begin to appear at negative $y$ values, indicating that they are falling into the cavity. The total number density continues to decline over time as droplets either exit the top of the computational domain or re-merge with the pool at the cavity bottom. This behavior is reflected in the $y$-profiles, which evolve from a single peak to two peaks, with one moving in the positive direction and the other in the negative direction over time.

\begin{figure}
\centering
\includegraphics[width=\textwidth]{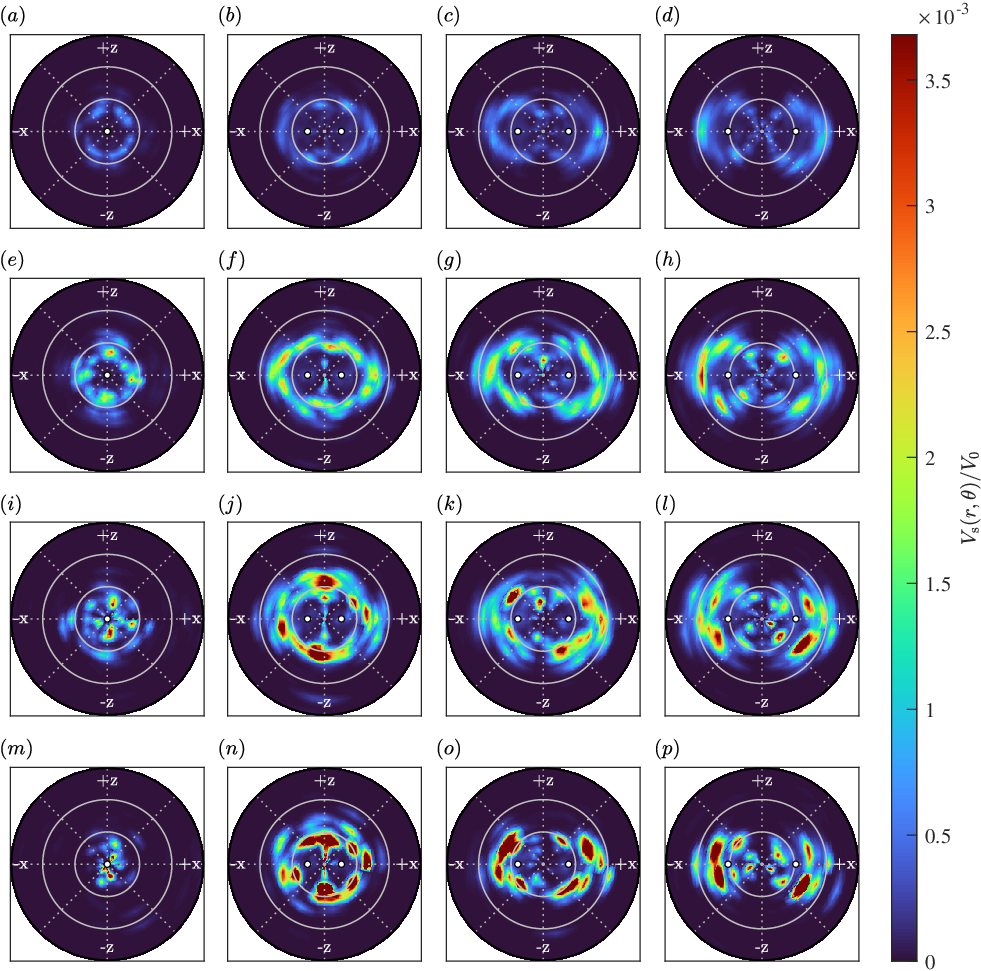}
\caption{\label{fig_polar} \textcolor{black}{Spatial distributions of the normalized airborne secondary droplet volume, $V_s(r,\theta)/V_0$. The columns from left to right correspond to cases SR, D2, D3, and D4, while the rows from top to bottom correspond to $t = 2.85$, 5.13, 9.12, and 11.97 ms. The maps are centered at $(x,z)=(0,0)$ and include only secondary droplets located above the undisturbed free surface. Each circular ring represents a radial distance of $2d_0$ from the origin, and the white markers denote the initial impact center locations.}}
\end{figure}

\textcolor{black}{To quantify the spatial spread of the ejected droplets, we plot the normalized secondary droplet volume $V_s/V_0$ (see figure~\ref{fig_polar}), for cases SR, D2, D3, and D4 from $t=2.85$ to $11.97~\mathrm{ms}$. Here $V_s$ is the total secondary droplet volume within each radial and azimuthal bin and $V_0$ is the initial raindrop volume. The influence of inter-raindrop interaction becomes most apparent at later times. In case SR, crown convergence toward the impact center causes the ligaments to merge into a single upward jet, limiting the lateral spread of airborne liquid (figure~\ref{fig_polar}c, e, and m). In contrast, the interacting crowns in the two-raindrop cases suppress this radial convergence, allowing secondary droplets to be distributed over a broader region.  Finally, the high-volume regions observed at later times (figure~\ref{fig_polar}n, o, and p) should not be interpreted as an increase in droplet number. Instead, they are associated with larger secondary droplets carrying greater individual volume.}

\subsubsection{Droplet re-merging dynamics}
\begin{figure}
\centering
\includegraphics[width=0.9\textwidth]{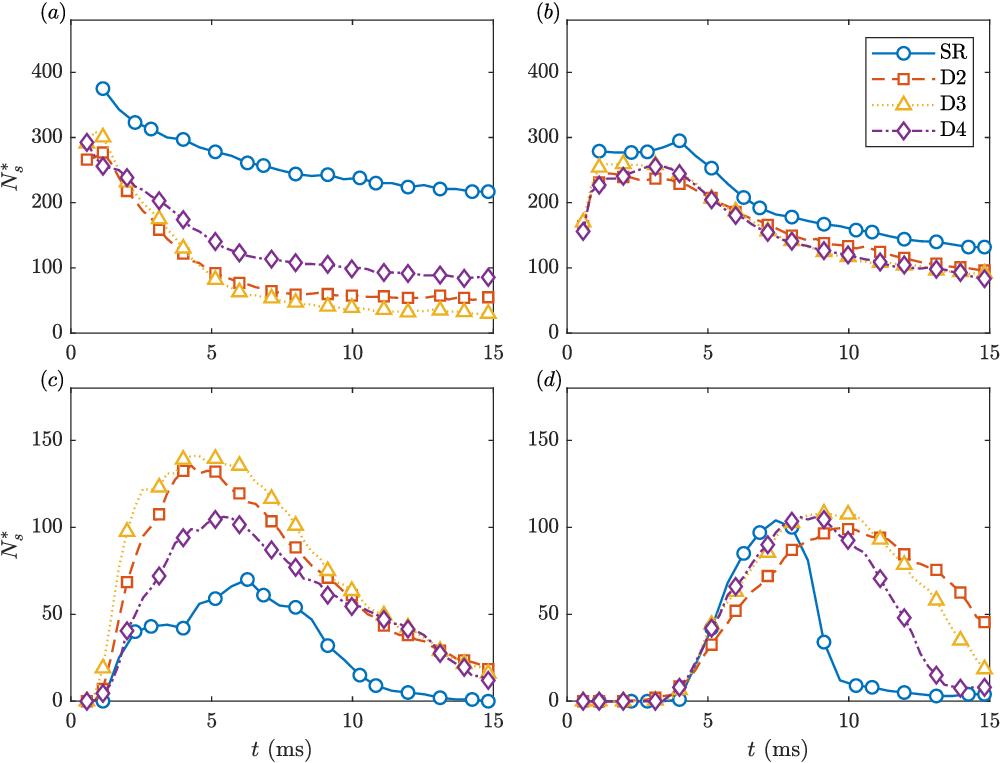}
\caption{\label{fig_int_compare}Time history of the normalized number of droplets, $N_s^*$, (a, b) above and (c, d) below $y=0$. Here, the droplet radii are (a, c) $r_s < 0.05\ $mm and (b, d) $0.05<r_s<0.1\ $mm, respectively.}
\end{figure}

\textcolor{black}{Here, we examine how inter-raindrop interaction affects the re-merging behavior of secondary droplets and separately plot $N_s^*$ above and below the unperturbed free surface, $y=0$, for droplets with $r_s<0.05~\mathrm{mm}$ and $0.05<r_s<0.1~\mathrm{mm}$ in figure~\ref{fig_int_compare}. Droplets below $y=0$ are interpreted as droplets that have entered the cavity, and the decrease of their population is used to assess their re-merging with the pool.}

For the droplets with $r_s < 0.05\,\mathrm{mm}$, their numbers remain largely unchanged in case SR until 15~ms (figure~\ref{fig_tot_size}a), whereas they gradually vanish in cases D2 and D3 (figure~\ref{fig_tot_size}b and~\ref{fig_tot_size}c). For case D4, the number follows a trend similar to case SR but with slightly fewer droplets overall (figure~\ref{fig_tot_size}d). As shown in figure~\ref{fig_int_compare}(a) and~\ref{fig_int_compare}(c), secondary droplets appear below $y=0$ as early as 1~ms, consistent with their early production. Relative to case SR, the two-raindrop cases show fewer droplets remaining above $y=0$ and more below $y=0$, \textcolor{black}{indicating that small droplets are more readily transported into the cavity in the two-raindrop cases, suggesting that the inter-raindrop interactions promote their downward transport.}

For droplets with $0.05<r_s<0.1\,\mathrm{mm}$, $N_s^*$ decreases drastically near 9~ms for case SR (figure~\ref{fig_tot_size}a), a feature absent in the two-raindrop cases. Their values above and below $y=0$ are plotted in figure~\ref{fig_int_compare}(b) and~\ref{fig_int_compare}(d), respectively. As shown in figure~\ref{fig_int_compare}(d), droplets appear below the surface later than for $r_s<0.05\,\mathrm{mm}$ (figure~\ref{fig_int_compare}c) at around 4~ms, reflecting their later production.  After about 8~ms, the values of $N_s$ below the surface begin to diverge, with a rapid decline first in case SR, followed by cases D4, D3, and D2. Hence, a similar number of secondary droplets enter the cavity across cases, but they re-merge with the pool at different rates. \textcolor{black}{The order of this decline shows that the re-merging of these droplets becomes progressively slower as the inter-raindrop spacing decreases.}

\begin{figure}
\centering
\includegraphics[width=0.9\textwidth]{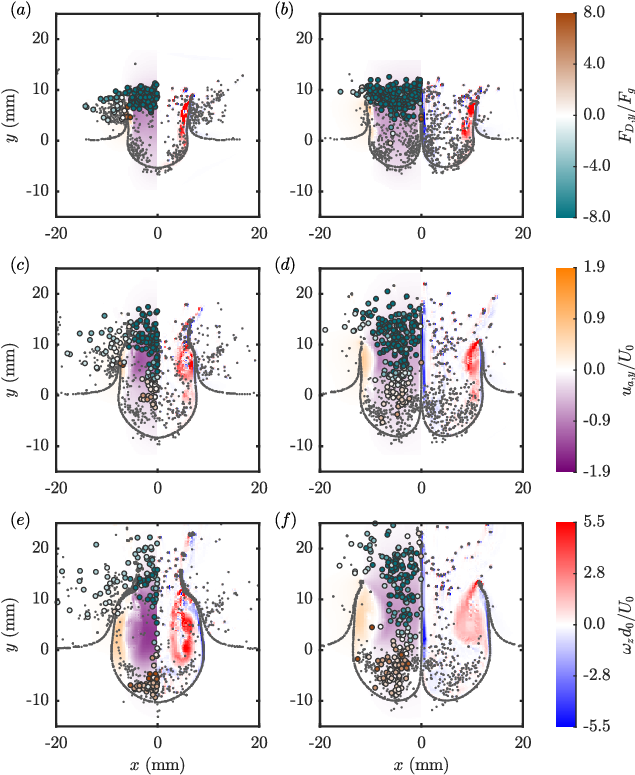}
\caption{\label{fig_SV_D2_flow_field}\textcolor{black}{Normalized vertical velocity and vorticity fields at (a,b) $t=2.28~\mathrm{ms}$, (c,d) $4.56~\mathrm{ms}$, and (e,f) $7.41~\mathrm{ms}$ for cases SR (a,c,e) and D2 (b,d,f) in the vertical plane of $z=0$. The left half of each plot shows the normalized vertical air velocity, while the right half shows the normalized vorticity. The positions of secondary droplets with $0.05<r_s<0.1~\mathrm{mm}$ are overlaid and colored by their normalized vertical drag. The air--water interface is denoted by grey lines. }}
\end{figure}

\textcolor{black}{The normalized vertical air velocity field, $u_{a,y}/U_0$, and normalized vorticity field, $\omega_z d_0/U_0$, are first examined inside the cavities to clarify how inter-raindrop interaction shapes the re-merging dynamics discussed above. In case SR (figure~\ref{fig_SV_D2_flow_field}a, c, and e), crown formation is accompanied by vortex formation near the crown rim as the inflowing air detaches, which induces a low-pressure region and a converging crown that further strengthens the downward airflow. This qualitatively agrees with the mechanism discussed by~\citet{wang2025splashing}. In contrast, in case D2 (figure~\ref{fig_SV_D2_flow_field}b, d, and f), the central film blocks air from entering the inner side of each cavity, reducing the development of the downward airflow.}

\textcolor{black}{We then compute the normalized vertical drag experienced by the droplets, $F_{D,y}/F_g$, where $F_g=(\rho_w-\rho_a)V_sg$ is the gravitational force, with $V_s=(4/3)\pi r_s^3$ being the effective droplet volume. The vertical drag is given by
$F_{D,y}=-(\pi r_s^2/2)\rho_a C_D |\boldsymbol{v}_s-\boldsymbol{u}_a|(v_{s,y}-u_{a,y})$,
where $\boldsymbol{v}_s$ is the droplet velocity, $\boldsymbol{u}_a$ is the local air velocity, and $C_D$ is estimated using the Schiller--Naumann correlation~\citep{schiller1933drag}. Positive values of $F_{D,y}/F_g$ indicate upward drag from the surrounding air, whereas negative values indicate downward drag.}

\begin{figure}
\centering
\includegraphics[width=\textwidth]{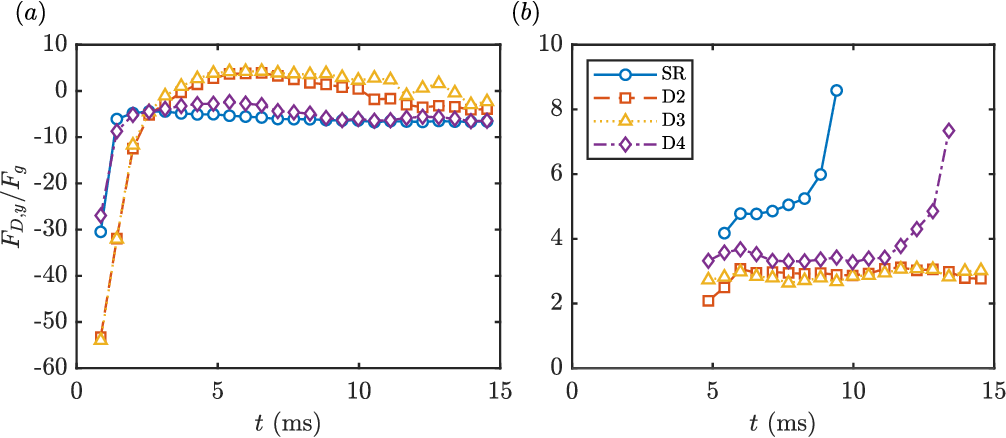}
\caption{\label{fig_drag}\textcolor{black}{Time histories of the median normalized vertical drag, $F_{D,y}/F_g$, for secondary droplets with (a) $r_s<0.05~\mathrm{mm}$ over the entire domain and (b) $0.05<r_s<0.1~\mathrm{mm}$ located below the undisturbed free surface, $y=0$.}}
\end{figure}

\textcolor{black}{Figure~\ref{fig_drag} shows the time history of the median $F_{D,y}/F_g$ for droplets of different sizes. For smaller droplets with $r_s<0.05~\mathrm{mm}$, the median $F_{D,y}/F_g$ has large negative values at early times (figure~\ref{fig_drag}a), which is a consequence of their rapid upward ejection and small gravitational force. Later, the drag force becomes positive or nearly neutral in cases D2 and D3, as the droplets are captured into the cavity, whereas in cases SR and D4, the force remains mostly negative. This is consistent with our previous observations that a larger fraction of droplets remain outside the cavity in these two cases (figure~\ref{fig_int_compare}a, c).}

\textcolor{black}{For larger droplets with $0.05<r_s<0.1~\mathrm{mm}$, we examine those located below the undisturbed free surface ($y=0$) (figure~\ref{fig_drag}b). The median drag force increases much more abruptly in case SR than in cases D2 and D3, while case D4 exhibits a delayed drag increase. This behavior can be explained by the drag forces acting on these larger droplets, as plotted in figure~\ref{fig_SV_D2_flow_field}. In case SR, most droplets have fallen below $y=0$ by $7.4~\mathrm{ms}$, and no additional droplets can enter the cavity after its closure (figure~\ref{fig_SV_D2_flow_field}e). Consequently, the median drag force increases sharply as the droplets decelerate while approaching the cavity bottom. In case D2, however, the capture of droplets by the cavity (figure~\ref{fig_SV_D2_flow_field}f) continues for a much longer duration, resulting in the relatively steady median drag force observed in figure~\ref{fig_drag}(b). We note that this overall trend is also consistent with the remerging behavior discussed in figure~\ref{fig_int_compare}(d).}

\section{Conclusion}
\label{sec:conclusion}
In this study, the fate of secondary droplets produced by the impact of raindrops interacting with a deep liquid pool has been examined through multi-phase DNS, accounting for varying surface tension, raindrop diameter, and raindrop interactions. At an early stage of the impact, the secondary droplet size distribution has been found to adhere to the scaling law of $N_d(r_s)\propto Q(\sigma/\rho_w)^{-1/2}r_s^{-5/2}$. Simulation data confirm this new scaling law, as well as the number density dependence on the surface tension and raindrop diameters. 

The raindrop interaction has been examined to understand the spatial and temporal distributions of secondary droplets. It is found that the central film formed by the merging of adjacent rising crowns plays a key role in shaping the overall impact morphology. As the inter-raindrop distance increases, the film forms later and ruptures less violently, leading to weaker disintegration, reduced bulge formation, and a shallower cavity. The impact of raindrop interaction on droplet distribution has been found to be size sensitive: smaller droplets with $r_s<0.05\,\mathrm{mm}$ are increasingly ejected upward rather than guided into the cavity as inter-raindrop distance grows, whereas those with $0.05 < r_s < 0.1\,\mathrm{mm}$ disappear more quickly into the pool in the single-raindrop case than in the two-raindrop cases. \textcolor{black}{The aerodynamic forcing analysis shows that this difference is associated with the cavity airflow, as the central film in the two-raindrop cases weakens the downward airflow and reduces the downward acceleration of droplets.}

Having examined the secondary droplet dynamics from high-speed raindrops in detail and established their behavior during the interactions between adjacent raindrops, the present work provides a basis for several future directions in modeling oceanic precipitation. Higher maximum refinement levels and finer temporal resolution will allow droplets from the prompt splash to be fully captured and analyzed. Expanding the computational domain would enable simulations to track the maximum travel distances of secondary droplets, providing additional insight into their influence on air--sea interactions. \textcolor{black}{In addition, although the present results provide a statistical scaling for the secondary-droplet population, a systematic investigation of the breakup mechanism remains an important direction for future work, since crown-rim instability, ligament formation, capillary breakup, and unsteady airflow may all contribute at these high impact speeds. The agreement of the two-raindrop cases with the proposed scaling suggests that the scaling may remain relevant for more general multi-raindrop impacts, at least locally where neighboring drop interactions dominate. A natural extension of the simulations presented here would therefore involve a larger number of adjacent raindrops, similar to the laboratory rainfall experiments~\citep[e.g.][]{liu2024experimental}. We note, however, that this will depend on continued advancements in computing power.}


\backsection[Acknowledgements]{\color{black}{The anonymous referees are gratefully acknowledged for their comments and suggestions.}}

\backsection[Funding]{\color{black}{X. H. is supported by UC San Diego.} }

\backsection[Declaration of interests]{The authors report no conflict of interest.}

\backsection[Data availability statement]{\color{black}{The data that support the findings of this study are openly available in [repository name] at http://doi.org/[doi], reference number [reference number].}}

\backsection[Author ORCIDs]{H.-H. Kuo, https://orcid.org/0009-0007-6874-5497; X. Hao, https://orcid.org/0000-0003-4898-1074}

\appendix
\section{Impact Geometry Extraction}
\label{app:impact_geometry}
\subsection{Automatic Cavity Geometry Extraction}
\begin{figure}
\centering
\includegraphics[width=0.8\textwidth]{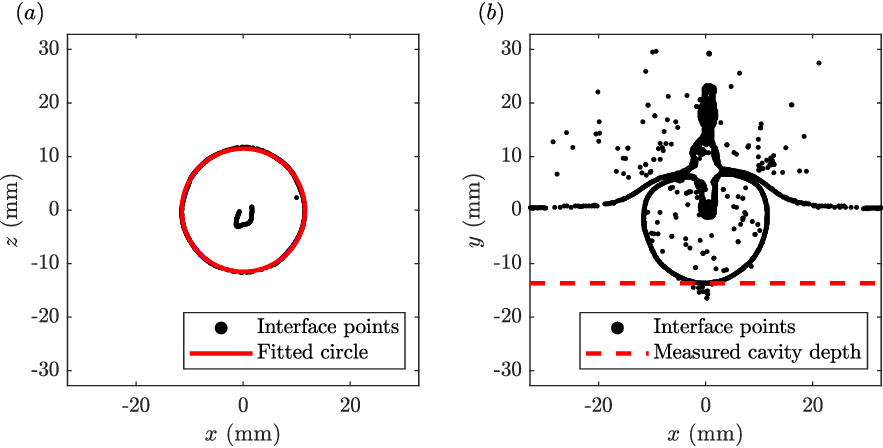}
\caption{\label{radfit}Example of (a) the circular fitting used to determine cavity radius and (b) the identified cavity depth location relative to the interface point at 24.51 ms.}
\end{figure}

To quantify the cavity radius, the coordinates of interface cells located near the undisturbed free surface (\(|y| < 0.041\)mm) are extracted at each time step. An initial least-squares circular fit is applied to these interface points in the \(x\text{--}z\) plane to estimate a preliminary center and estimated radius \(r_e\). The radial distance of each point from this estimated center is denoted by \(r\). To remove outliers, only points satisfying \(|r - r_e| < 2 \times \mathrm{median}(|r - r_e|)\) are retained. Next, a circular arc is fitted for each point using its five nearest neighbors in the \(x\text{--}z\) plane to estimate the local curvature. Points with curvature exceeding a predefined threshold are excluded from the fit. After these filtering steps, a final least-squares circular fit is performed on the remaining points. The radius of this final circle is taken as the cavity radius \(r_c\). The thresholds for radial deviation, curvature, and neighbor count are empirically selected by iteratively adjusting them to achieve good agreement between the fitted circle and the cavity interface. An example plot at $t = 24.51\ \mathrm{ms}$ is shown in figure~\ref{radfit}(a).

The cavity depth is determined by first identifying interface cells in the domain and computing their curvature. Only those with negative curvature and magnitude below a threshold of $0.122\ \mathrm{mm}^{-1}$ are retained. The five vertical neighboring cells above each candidate are then examined; if all five are fully air, the candidate is retained; otherwise, it is discarded. These two criteria prevent the algorithm from incorrectly identifying interface cells belonging to isolated air bubbles beneath the cavity. The minimum \(y\)-coordinate among the remaining candidates is then taken as the cavity depth. An example plot at $t = 24.51\ \mathrm{ms}$ is shown in figure~\ref{radfit}(b).

\subsection{Automatic Crown Rim Geometry Extraction}
\begin{figure}
\centering
\includegraphics[width=0.8\textwidth]{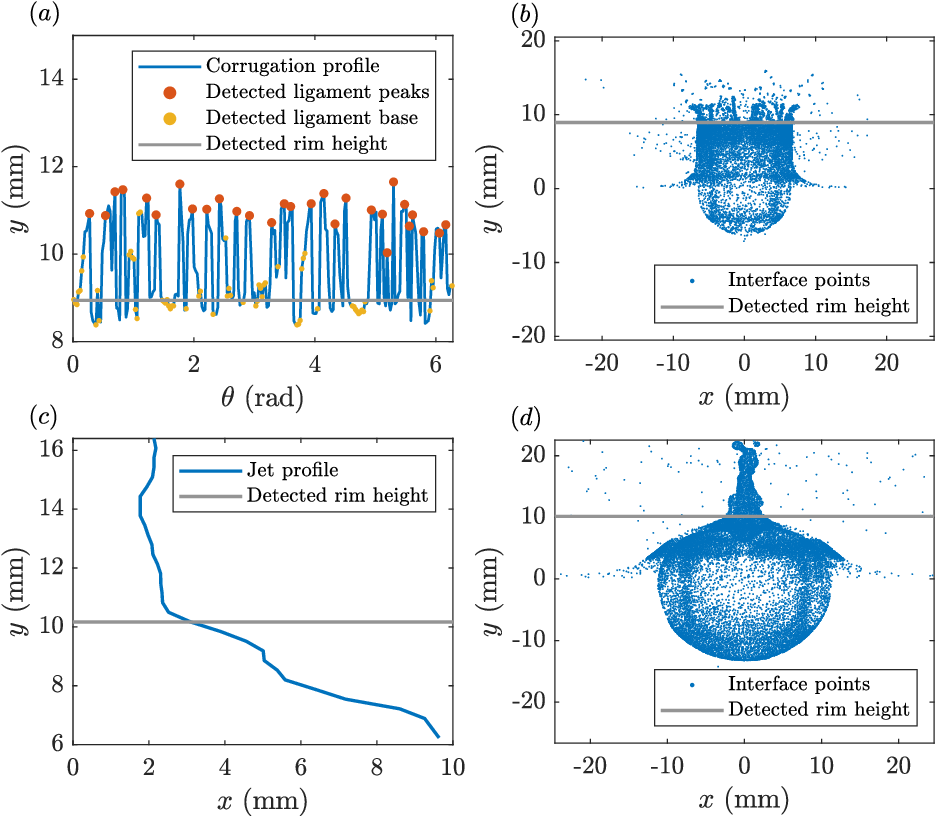}
\caption{\label{figraddetect}
{\color{black}(a) Azimuthal profile of the crown rim with ligament peaks and bases, and (b) air--water interface points at $t = 2.85\ \mathrm{ms}$. (c) Radial width of the jet and rim, and (d) interface points at $t = 17.67\ \mathrm{ms}$. Also plotted are the detected rim heights.}}
\end{figure}

{\color{black}The crown rim height and radius are determined using two different procedures because the crown geometry changes during the impact process. In the early stage, the individual ligaments around the rim remain distinct. The interface points around the crown rim are first used to extract the azimuthal variation of the rim profile (see an example in figure~\ref{figraddetect}a). The crown rim height is then defined as the median height of the locations of all ligament bases.  Figure~\ref{figraddetect}(b) shows the detected rim height relative to the air--water interface points. At the rim height, the radial distances of the interface points from the crown center are computed in the \(x\text{--}z\) plane to estimate the rim radius. 

In the later stage, after around \(11\ \mathrm{ms}\), the rim converges radially, and the ligaments merge into a single liquid jet.  The radial width of the interface is first extracted as a function of height, and the rim height is selected at the transition between the crown and the jet-like structure (figure~\ref{figraddetect}c). The rim radius is obtained directly from the radial-width profile at the rim height. Figure~\ref{figraddetect}(d) shows the detected rim height, the air--water interface points, and the jet-like structure.}

\begin{figure}
\centering
\includegraphics[width=\textwidth]{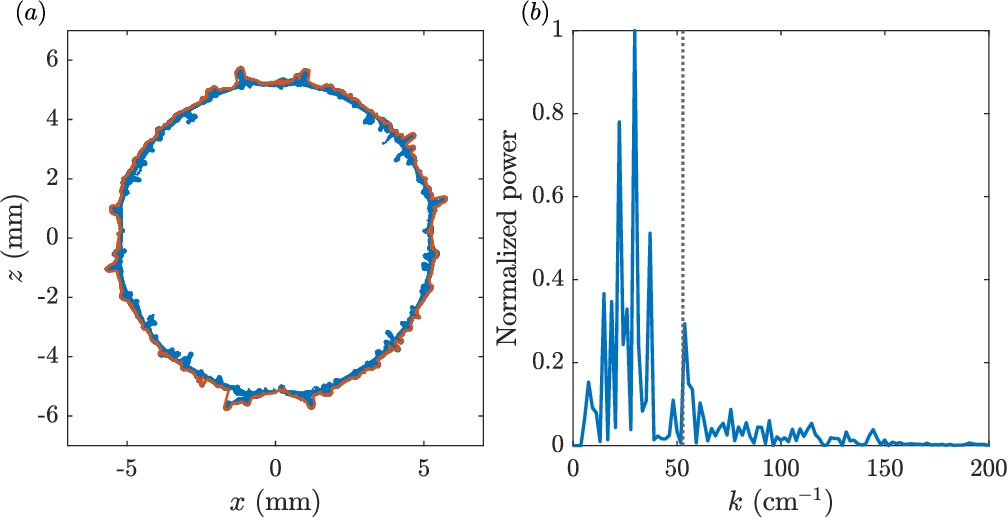}
\caption{\label{rimprof}(a) Air--water interface of the crown in case SR at $y=2.25$ mm above the unperturbed free surface at $1.14$ ms. Here, the orange outline represents the extracted corrugation profile around the rim. (b) Power spectrum of the rim corrugation profile, i.e., the orange curve in (a), normalized by its maximum value. The vertical black dotted line marks the theoretical wavenumber with the maximum growth rate predicted by the Rayleigh--Plateau instability model.}
\end{figure}

{\color{black}
\section{Characteristics of Crown Corrugation and Stability Analysis}
\label{app:crown_corrugation}
Following~\citet{zhang2010wavelength}, we test the feasibility of the Rayleigh--Plateau instability in predicting the growth of perturbations on the crown rim and the subsequent production of secondary droplets from these perturbations. In this framework, the local rim radius is written as
\begin{equation}
r(\theta,t)=r_0(t)+\epsilon(\theta,t),
\end{equation}
where $r_0(t)$ is the mean rim thickness, $\theta$ is the azimuthal angle, and $\epsilon(\theta,t)$ is a small perturbation. The initial perturbation is decomposed into azimuthal Fourier modes,
\begin{equation}
\epsilon(\theta,0)=\sum_{m=-N}^{N} a_m e^{im\theta},
\end{equation}
where $m$ is the azimuthal mode number and $a_m$ is the initial amplitude of each mode. The physical wavenumber of mode $m$ is
\begin{equation}
k_m(t)=\frac{m}{R(t)},
\end{equation}
where $R(t)$ is the crown radius. The dimensionless wavenumber entering the Rayleigh--Plateau dispersion relation is therefore
\begin{equation}
x_m(t)=k_m(t)r_0(t)=\frac{m r_0(t)}{R(t)}.
\end{equation}

For each mode, the viscous Rayleigh--Plateau growth rate $\sigma_m$ is obtained from the classical stability result for a viscous cylindrical thread,
\begin{equation}
\begin{aligned}
&(2x^2)(x^2+y^2)\frac{I_1'(x)}{I_0(x)}
\left[
1-\frac{2xy}{x^2+y^2}
\frac{I_1(x)}{I_1(y)}
\frac{I_1'(y)}{I_1'(x)}
\right]
-x^4+y^4  \\
&\qquad =
\frac{\gamma r_0}{\rho \nu^2}
\frac{xI_1(x)}{I_0(x)}(1-x^2),
\end{aligned}
\end{equation}
where $y^2=x^2+{\sigma_m r_0^2}/{\nu}$, $\gamma$, $\rho$, and $\nu$ are the liquid surface tension, density, and kinematic viscosity, respectively, and $I_n\ (n=0,1)$ denotes the modified Bessel function of the first kind.

Because the crown rim expands in time, the perturbation amplitude is affected by both capillary growth and rim stretching. Following~\citet{eggers2008physics}, the amplitude of each mode evolves as
\begin{equation}
\frac{d\ln \epsilon_m}{dt}
=
-\frac{s}{2}
+
\sigma_m,
\qquad
s=\frac{\dot R}{R}.
\end{equation}
After integration, this gives
\begin{equation}
\epsilon_m(t)
=
a_m
\left(\frac{R(0)}{R(t)}\right)^{1/2}
\exp\left[
\int_0^t \sigma_m(t')\,dt'
\right].
\end{equation}
The theoretical rim profile is then reconstructed as
\begin{equation}\label{eq:Rayleigh_Plateau_theory_rim}
r(\theta,t)
=
r_0(t)
+
\left(\frac{R(0)}{R(t)}\right)^{1/2}
\sum_{m=-N}^{N}
a_m
\exp\left[
im\theta+
\int_0^t \sigma_m(t')\,dt'
\right].
\end{equation}

Using equation~\eqref{eq:Rayleigh_Plateau_theory_rim}, we compute the theoretical corrugation profile and its wavenumber spectrum with $R(t)$ and $r_0(t)$ extracted from the simulations. The theoretical dominant wavenumber is then identified from the mode at the spectrum peak. The measured spectrum is obtained independently from the simulated interface. The interface points near the crown are projected onto the $x$-$z$ plane to reconstruct the rim outline. The deviation of the extracted outer rim radius from its mean defines the measured corrugation profile $h$ along the rim (figure~\ref{rimprof}a) and the corresponding power spectrum (figure~\ref{rimprof}b).

As shown, the peak wavenumber obtained from the simulation does not coincide with the theoretical prediction by the Rayleigh--Plateau model. Similar comparisons from $t=1.14\ \mathrm{ms}$ to $t=4\ \mathrm{ms}$ show persistent deviations (results not shown). This discrepancy is likely related to the difference between the present impact regime and the crown-splash regime studied by~\citet{zhang2010wavelength}. Their experiments used silicone oil and moderate impact velocities, giving relatively small $Re\sim1000$ and $We\sim800$, and the crown corrugations evolved smoothly from small amplitude perturbations. In contrast, our simulations occur at much higher Reynolds and Weber numbers, i.e. $Re\sim30000$ and $We\sim3000$, where the rim deforms violently and the crown breakup is highly fragmented and irregular (figure~\ref{rimprof}a). Such high irregularity in the crown corrugation continues to be observed in all of our cases. Therefore, it is inappropriate to treat droplet production solely as the growth of rim corrugations through the Rayleigh--Plateau instability in the present impact regime.}

\bibliographystyle{jfm}
\bibliography{jfm} 
\end{document}